\documentclass[onecolumn]{aa}
\usepackage{graphicx}
\usepackage{txfonts}
\begin{document}
   \title{uvby-$\beta$ photometry of high-velocity and metal-poor stars}

   \subtitle{X. Stars of very low metal abundance:  observations, reddenings,
             metallicities, classifications, distances, and relative 
             ages\thanks{Based on observations collected at the H.L.~Johnson
             $1.5\,$m telescope at the Observatorio Astron\'omico Nacional at 
             San Pedro M\'artir, Baja California, M\'exico, and at the Danish 
             $1.5\,$m telescope, La Silla, Chile.}
	     }

   \author{W.J. Schuster\inst{1,2}, T.C. Beers\inst{3}, R. Michel\inst{1},
          P.E. Nissen\inst{4}, \and  G. Garc\'{\i}a\inst{1}
          }

   \offprints{W.J. Schuster}

   \institute{Observatorio Astron\'omico Nacional, UNAM, Apartado Postal
              877, Ensenada, B.C., M\'exico, C.P. 22800\\
              \email{schuster@astrosen.unam.mx, rmm@..., gabi@...}
	 \and
	      Instituto Nacional de Astrof\'{\i}sica, \'Optica y Electr\'onica, Luis
	      Enrique Erro No.~1, Tonantzintla, Puebla, M\'exico, C.P. 72840\\
	      \email{schuster@inaoep.mx}              
         \and
              Department of Physics and Astronomy, Michigan State University,
	      East Lansing, Michigan 48824, U.S.A.\\
              \email{beers@pa.msu.edu}
         \and
	      Department of Physics and Astronomy, University of Aarhus, DK-8000
	      Aarhus C, Denmark\\
	      \email{pen@phys.au.dk}
              }

   \date{Received October 15, 2003; accepted November 16, 2003}

   \abstract{$uvby$(--$\beta$) photometry has been obtained for an additional 411 very
   metal-poor stars selected from the HK survey, and used to derive basic parameters 
   such as interstellar reddenings, metallicities, photometric classifications, 
   distances, and relative ages.  Interstellar reddenings adopted from the Schlegel 
   et al.~(1998) maps agree well with those from the intrinsic-color calibration of 
   Schuster \& Nissen (1989).  [Fe/H] values are obtained from the CaII K line index 
   of the HK survey combined with the $uvby$ and $UBV$ photometry.  The 
   $c_{\rm 0},(b-y)_{\rm 0}$ diagram is seen to be very useful for classifying these 
   very metal-poor field stars into categories similar to those derived from globular 
   cluster color-magnitude diagrams; it is found that the
   HK survey has detected metal-poor candidates extending from the red-giant to the 
   blue-horizontal branch, and from the horizontal branch to subluminous stars.  
   Distances derived from $UBV$ photometry agree reasonably well with those from 
   $uvby$, considering the paucity of good calibrating stars and the extrapolations 
   required for the most metal-poor stars.  These very metal-poor stars are compared 
   to M92 in the $c_{\rm 0},(b-y)_{\rm 0}$ diagram, and evidence is seen for field stars 
   1--3 Gyrs younger than this globular cluster; uncertainties in the [Fe/H] scale for 
   M92 would only tend to increase this age difference, and significant reddening 
   uncertainties for M92 are unlikely but might decrease this difference.  The 
   significance of these younger very metal-poor stars is discussed in the context of 
   Galactic evolution, mentioning such possibilities as hierarchical 
   star-formation/mass-infall of very metal-poor material and/or accretion events 
   whereby this material has been acquired from other (dwarf) galaxies with different 
   formation and chemical-enrichment histories.

   }
   \authorrunning{W. J. Schuster et al.}
   \titlerunning{Very metal-poor stars}
   \maketitle
%

\section{Introduction}

Over the past two decades, our collective knowledge of the nature of the thick disk
and halo of the Galaxy has expanded enormously, due primarily to the impact of
several ongoing large-scale survey efforts carried out to detect and analyze
metal-poor stars. These include the HK survey of Beers and collaborators (Beers,
Preston, \& Shectman 1992; Beers 1999) and the Hamburg/ESO stellar survey of
Christlieb \& collaborators (Christlieb 2003), both of which select stars with
objective-prism techniques, and hence introduce no kinematic bias into their samples.
Such biases are present (and must be corrected for) in proper-motion selected
survey samples, such as the exhaustive previous studies of, e.g., Ryan \& Norris 
(1991) and Carney et al.~(1996). The prism-survey selected samples are hence
well suited for studies of the kinematics and dynamics of the old stellar
populations of the Milky Way, in particular because of the burgeoning databases
of proper motion information that are presently being assembled from a variety
of sources (e.g., UCAC2: Zacharias 2002; SPM: Girard et al.~2003). In order to
make optimal use of the proper motions for kinematic analyses, accurate stellar
classifications, and photometrically determined distances, are crucial.

The $uvby$--$\beta$ photometric system is particularly suited for the study of 
very-metal-poor (hereafter, VMP) F- and G-type stars, as has already been pointed 
out in Paper VIII by Schuster et al.~(1996; hereafter S96).  Briefly, 
intrinsic-color calibrations, $(b-y)_{\rm 0}$--$\beta$, exist which allow
accurate and precise, $\pm 0\fm01$, measures of interstellar reddening excesses,
$E(b-y)$, for individual field stars; such a calibration has been given by
Schuster \& Nissen (1989).  Photometric absolute magnitudes and distances can
be calibrated and used effectively, as shown in the papers by Olsen (1984) and 
Nissen \& Schuster (1991).  This photometric system has the great advantage that 
it permits us to obtain accurate stellar distances even for evolving main-sequence
and subgiant stars due to the gravity sensitivity of the $c_{\rm 0}$ index.  Also,
importantly, theoretical isochrones in the $M_{\rm bol}$, $T_{\rm eff}$ diagram
can be transformed to the $M_{\rm V}$, $(b-y)_{\rm 0}$ or $c_{\rm 0}$, 
$(b-y)_{\rm 0}$ diagrams for the estimation of relative and/or absolute ages
of evolving field stars which are near their respective turn-offs, and in 
several of the previous papers of this series the isochrones of VandenBerg et 
al.~have been used for such purposes, to study the Galactic halo population and 
to make comparative analyses between the relative ages of the halo and thick-disk
stellar populations.  Most recently the isochrones of Bergbusch \& VandenBerg
(2001) have been transformed to the $uvby$ photometric system using the
color--$T_{\rm eff}$ relations of Clem et al.~(2003).

Also, the $uvby$--$\beta$ photometry can provide basic stellar
atmospheric parameters as a prelude to detailed chemical abundance studies
making use of high-resolution spectroscopy and model atmospheres.  Several
empirical calibrations already exist in the literature for the conversion
of $(b-y)_{\rm 0}$ or H$\beta$ to $T_{\rm eff}$; these calibrations include 
appropriate metallicity dependences.  Index diagrams, such as 
$c_{\rm 0}$, $(b-y)_{\rm 0}$, or the reddening-free $[c_{\rm 1}]$, $[m_{\rm 1}]$, 
or $[c_{\rm 1}]$, $\beta$, allow the classification of field stars according 
to their evolutionary status, permiting us to estimate the stellar surface 
gravities, also for input into the model-atmosphere analyses.

In this paper, $uvby$--$\beta$ photometry is presented for an additional 411
VMP stars from the HK survey, providing a total database of such photometry for
497 VMP stars, when combined with the data of S96.  For the present sample the 
stars have been selected with [Fe/H] $\la -1.5$, and 243 were observed in M\'exico 
using classical photometric (photoelectric) techniques and 177 in Chile using 
DFOSC (CCD) techniques.  In Sect.~2 the observing and reduction techniques
are described briefly, the catalogues of new $uvby$--$\beta$ data presented,
and the $V$ magnitudes and $(b-y)$ colors from the $uvby$ observations compared 
to magnitudes and $(B-V)$ from the HK survey.  In Sect.~3, the photometry is
dereddened using a modification of the Schlegel et al.~(1998) reddening maps
and also the intrinsic-color calibration of Schuster \& Nissen (1989);
reddenings from the two methods, $E(B-V)$ and $E(b-y)$, are compared.  In
Sect.~4, [Fe/H] values are derived for the VMP stars using the techniques
developed in the HK survey, and probable carbon-enhanced stars are identified
based on a comparison of the GP and KP indices.  Photometric classifications
are derived for the VMP stars in Sect.~5 using the $c_{\rm 0}$, $(b-y)_{\rm 0}$ 
diagram.  Stars are found covering a wide range of stellar types from the 
horizontal branch (HB) to subluminous stars (SL), and from the
red giant stars (RG) to the blue horizontal branch (BHB), and other categories
include main-sequence (MS), turn-off (TO), subgiant (SG), blue-straggler (BS),
and red-horizontal-branch-asymptotic-giant-branch (RHB-AGB) stars. Possible 
abundance anomalies for some VMP stars have been identified from the
$uvby$ photometric indices and diagrams, such as the $[c_{\rm 1}]$, $[m_{\rm 1}]$; 
for example, ten probable Am stars have been found and also a 
number of possible AGB stars with unusual chemical abundance ratios or 
binary companions.  Distance estimates are made for the VMP stars in Sect.~6 
using $uvby$ photometry plus various methods and new calibrations, and
also using the $UBV$ photometry and techniques developed in the HK survey.
Comparisons of these photometric distances show reasonably good agreement,
considering the paucity of calibrating stars and extrapolations required for
the more VMP stars.  In Sect.~7, the VMP field stars are compared to the
globular cluster M92 in the $c_{\rm 0}$, $(b-y)_{\rm 0}$ diagram, using the
isochrones of Bergbusch \& VandenBerg (2001), as transformed to $uvby$ by
Clem et al.~(2003), to interpolate relative and absolute ages.  A number of VMP
stars apparently 1--3 Gyrs younger than M92 are noted, and their importance
for understanding the formation and evolution of the Galactic halo discussed.

\section{Photometric observations of the very metal-poor stars}

\subsection{Selection of the stars}


The VMP stars described herein were selected from
two primary sources. The first set of 194 stars is a subset of the published
catalogues of Beers, Preston, \& Shectman (1985; BPSI) and Beers, Preston, \&
Shectman (1992; BPSII), using the criterion [Fe/H]$_c \le -1.5$, where [Fe/H]$_c$
is the corrected spectroscopic metallicity estimate derived by BPSII based on a
calibration of the strength of the CaII K index KP as a function of measured or
inferred $(B-V) _0$ color. This set includes, primarily, stars at or near the
main-sequence turnoff and warmer subgiants. The second set of 303 stars were
{\it candidate} VMP stars selected from visual inspection of medium-resolution
spectroscopy obtained during the course of the HK survey follow-up at a number
of observatories, and includes stars covering a larger range of effective
temperatures and luminosities. Since this second subset was selected prior to
obtaining estimates of metallicity, it includes a larger fraction of stars
exceeding [Fe/H] $= -1.5$ than the BPSI and BPSII subsample. The full set of
HK-survey medium-resolution spectroscopic results will appear in a series of
papers in preparation. 

\subsection{Observation and reduction techniques}

The $uvby$--$\beta$ data presented here for the VMP stars were taken using
$1.5\,$m telescopes and two different types of instrumentation.  
The data of Table 1 were taken during ten observing runs from September 1995 
through November 2000 at the H.L. Johnson $1.5\,$m telescope at the San Pedro 
M\'artir Observatory, Baja California, M\'exico (hereafter SPM), and the $uvby$ 
data of Table 2 during three runs from October 1998 through September 2000 at 
the Danish $1.5\,$m at the European Southern Observatory, La Silla, Chile 
(hereafter La Silla).  For the SPM observing a six-channel $uvby$--$\beta$ 
photoelectric photometer was used, the same as for the Schuster \& Nissen 
(1988; hereafter SN) and Schuster et al.~(1993; hereafter SPC) catalogues 
and for the $uvby$--$\beta$ data of VMP stars by S96.  For the La Silla 
observing the DFOSC has been used with a CCD detector, as described by Brewer 
\& Storm (1999).

The $uvby$--$\beta$ data presented here for the VMP stars in Table 1 were taken 
and reduced using techniques very nearly the same as for SN, SPC, and S96.
One is referred to these previous papers for more details; what 
follows is a brief outline of the more important points of the observational 
techniques.  The four-channel $uvby$ section of the SPM photometer is really a
spectrograph-photometer which employs exit slots and optical interference
filters to define the bandpasses.  The grating angle of this 
spectrograph-photometer was adjusted at the beginning of each 
observing run to position the spectra on the exit slots to within 
about $\pm 1{\AA}$.  Whenever possible,
extinction-star observations were made nightly over an air-mass interval of
at least 0.8 (see Schuster \& Parrao 2001), and spaced throughout each night
several ``drift'' stars were observed symmetrically with respect to the local
meridian, and with these observations the atmospheric extinction coefficients
and time dependences of the night corrections could be obtained for each of
the nights of observation (see Gr{\o}nbech et al.~1976).  Finding charts
were employed at the SPM and La Silla telescopes to identify all of 
the stars from the HK survey.  For previous studies, such as S96,
the program stars were observed at SPM to at least 50,000 counts in all 
four channels of $uvby$ and to at least 30,000 counts for the two channels 
of H$\beta$; here, the fainter program stars at SPM ($V \ga 14\fm5$) were 
exposed to only at least $\approx 30,000$ counts in all four channels of $uvby$, 
and H$\beta$ was observed only for the brighter program stars ($V \la 14\fm0$) 
and to only $\approx 20,000$ counts in both channels at $V \approx 14\fm0$.  
For all program stars the sky was measured until its contributing error was 
equal to or less than the error of the stellar count.  At SPM an attempt 
was made to obtain three or more independent $uvby$ observations for each 
of the program stars.

The $uvby$ observations for the VMP stars of Table 2 were taken using the
C1W7 CCD (LORAL/LESSER backside illuminated chip) with 15 micron pixels,
and a ESO $uvby$ filter set (Nos. 715, 716, 717, and 718).  A more or less
clean and uniform part of the chip was selected for the observations, and
the Midas routine ``point'' was used to position the stars near the center
of this area; for most nights the RMS positioning error was better than $\pm 2$ 
pixels, except for the more windy nights when it was $\ga \pm 2$ pixels.
Since single stars were being observed, an area of only $250 \times 250$ 
pixels was read out around the center of this ``point'' routine.  In this 
way the observations could be read out rapidly and the filters cycled more 
quickly:  $ybvu$ or $ybvuuvby$, with a bias taken after each
cycle.  Also, by reading $250 \times 250$ pixels, four or more sky flats could be
obtained in each filter-band during both the evening and morning twilights.
By always centering the stars very nearly at the same place on the CCD, we
could avoid major cosmetic defects, and also several problems of flat fielding,
such as variations in the dispersed light.  Whenever possible, atmospheric
extinction observations were made nightly over an air-mass interval of at 
least 0.8.  Extra biases were measured at the beginning and (sometimes) end
of the nights, and a few 800s (or longer) dark measures were made during the 
observing runs (800s being the longest stellar integration). In general we 
attempted to obtain at least 22,000 ADUs in all bands for the program stars, 
corresponding to about 30,000e$^-$, and to obtain at least two independent 
observations for the program stars; this latter criterion was not 
accomplished for 86 of the La Silla program stars due mainly to poor 
photometric conditions during the last observing run.

For the CCD data from La Silla the IRAF package was used for the image 
reduction.  All the images were bias, dark, and flat-field corrected 
employing the usual routines. The program and standard stars of this study 
were identified in all the fields and their centroids calculated.  For 
each observed night, the FWHMs of all objects were averaged and from 
this average three different apertures from 3 to 6 times the $<$FWHM$>$ were 
tested.  The PHOT routine was then used for extracting the instrumental 
magnitudes of all objects in the different filters.  These instrumental 
values were then fed into the reduction programs of T. Andersen, described 
below, and reduced in the usual fashion.  That extraction aperture which 
gave the smallest instrumental and transformation errors was then retained 
for the reduction of the final program-star standard photometry.

As for the SN, SPC, and S96 catalogues, all of these data reductions
were carried out following the precepts of Gr{\o}nbech et al.~(1976)
using computer programs kindly loaned by T. Andersen.  At SPM the 
$uvby$--$\beta$ standard stars observed were taken from the same lists as 
for the previous catalogues; these are mostly secondary standards from 
the catalogues of Olsen (1983,1984).  A few of the more metal-poor
stars from Olsen and from the SN catalogue (such as HD2796, HD84937, 
HD140283, HD195363, BD$-17$:0267, and CD$-24$:1782) were observed
often to be used as standard stars and to check for consistency.  At La Silla 
the standard-star list was derived from stars with $V \la 10\fm5$ from
SN, from S96, and from a 1998 version of our Table 1.  The reduction programs 
create a single instrumental photometric system for each observing run, 
including nightly atmospheric extinctions and night corrections with 
linear time dependences.  Then transformation equations from the 
instrumental to the standard systems of $V$, $(b$--$y)$, $m_{\rm 1}$, $c_{\rm 1}$,
and $\beta$ are obtained using all standard stars observed during that
observing period.  The equations for the transformation to the standard
$uvby$--$\beta$ system are the linear ones of Crawford \& Barnes (1970)
and of Crawford \& Mander (1966).  Small linear terms in $(b-y)$ 
are included in the standard transformation equations for $m_{\rm 1}$ 
and $c_{\rm 1}$ to correct for bandwidth effects in the $v$
filter.  Our $y$ measures have been transformed onto the V system of Johnson 
et al.~(1966).  For the S96 catalogue we had selected a more homogenous 
program list of VMP stars with [Fe/H]$_{\rm c} \leq -2.50$ and with the 
selected stars restricted mainly to the bluer ``TO'' types (turn-off star 
candidates) with a few ``SG'' types (subgiant candidates); a few in fact 
turned out to be horizontal-branch stars.  For the present catalogues all 
types of stars from the HK survey were thrown into the program-star observing
lists while the standard-star lists were extended only slightly to include
a few horizontal-branch stars and a few red subgiant/giants.  For this
reason the transformation equations to the standard system had to be
extrapolated for some of the more extreme stars, as seen below in the 
$c_{\rm 0}$,$(b$--$y)_{\rm 0}$ diagram of Fig.~6.  The standard photometry
of the BHB, SL--BHB, BS, SL, and some of the HB and BS--TO stars results 
from such an extrapolation.  For example, for each of the observing runs 
at La Silla, approximately 36 $uvby$ standard stars were observed in the 
following ranges:  $0\fm305 \leq (b$--$y) \leq 0\fm796$, 
$0\fm031 \leq m_{\rm 1} \leq 0\fm576$, and $0\fm123 \leq c_{\rm 1} \leq 0\fm726$.
These limits can be compared to the range of values displayed in Fig.~6, 
which has been dereddened.

The $uvby$ photometry from SPM is of higher quality than the $uvby$ data from
La Silla, due in part to differences in the instrumentation and in part due to
differing photometric qualities of the nights observed.  For the SPM
photometer the measures in the four bands are taken simultaneously, and so
several instrumental and atmospheric effects cancel out to large degree,
such as those due to atmospheric extinction and seeing.  The La Silla $uvby$
data were observed with the DFOSC, sequentially, and in general the nights
observed at La Silla were not of the same high photometric quality as those 
of SPM.  For the SPM $uvby$--$\beta$ data of Table 1, typical (median) values 
for the standard deviations of a single observation are $\pm0\fm008$, $0\fm007$,
$0\fm010$, $0\fm013$, and $0\fm014$ for $V$, $(b$--$y)$, $m_{\rm 1}$, $c_{\rm 1}$,
and $\beta$, respectively.  For the La Silla $uvby$ data of Table 2, the typical 
standard deviations of a single observation are $\pm0\fm009$, $0\fm015$, 
$0\fm021$, and $0\fm023$ for $V$, $(b$--$y)$, $m_{\rm 1}$, and $c_{\rm 1}$,
respectively.

\subsection{The catalogues of observations}

Table 1 presents the $uvby$--$\beta$ catalogue for the 243 VMP stars observed
at SPM; 156 of these stars have measured H$\beta$ values, mostly those with $V 
\la 14\fm0$.  Column 1 lists the stellar identifications according to the 
nomenclature of BPSI and BPSII, Col.~2 the $V$ magnitude on the 
standard Johnson $UBV$ system, and Cols.~3--5, and 7, the indices 
$(b$--$y)$, $m_{\rm 1}$, $c_{\rm 1}$, and $\beta$, respectively, on the
standard systems of Olsen (1983, 1984), which are essentially the systems
of Crawford \& Barnes (1970) and Crawford \& Mander (1966) but with a
careful extension to metal-poor stars and with north-south systematic
differences corrected.  Columns 6 and 8 give $N_u$ and $N_{\beta}$, the
total numbers of independent $uvby$ and $\beta$ observations, respectively.
Stars marked with a ``$++$'' in the ``Notes" column are red subgiant/giants,
($(b$--$y) \ga 0\fm5$ and $c_{\rm 1} \ga 0\fm35$); as discussed in SN,
the $m_{\rm 1}$ and $c_{\rm 1}$ values of these stars may be less accurate.

Column 9 of Table 1 lists notes for the VMP stars taken during the
observations or during the data reductions and analyses.  For example,
15621--051 shows indications of photometric variability in its
$uvby$--$\beta$ data, below is classified as a horizontal branch
(``HB'') star, and so may very well be a VMP RR-Lyrae-type star.  The star 
16033--081 is one of the red subgiant/giant stars mentioned above.  The 
photometry of 16549--043 was contaminated by a faint nearby star which was 
also included in the photometer's diaphragm during the observations.  The 
VMP star 17583--067 was offset in the photometer's diaphragm to exclude a 
fainter nearby star; since the bandpasses of the SPM photometer are mainly 
filter-defined, this small offset should produce negligible errors for the 
indices.  The star 22955--032 was observed in two different ways:  for two 
nights with poorer seeing, with a fainter nearby star also in the 
photometer's diaphragm, and for one night with good seeing, offset with this 
nearby star excluded; the photometric values with $N_u = 3$ include all 
three measures, those with $N_u = 2$ only the observations with the fainter 
star included, and $N_u = 1$ with this fainter star excluded.

For those stars noted as photometric variables (``V''), all eight, 15621--051,
16089--086, 16541--052, 16557--063, 17136--014, 17435--003, 22872--010, and 30320--075,
are classified as ``HB'', horizontal-branch, in the photometric classifications
to follow, and these are all good candidates for VMP RR-Lyrae-type stars.  Of the
six stars marked as possible photometric variables (``V?''), two more are
classified below as ``HB'' (16089--042 and 17570--011), one as ``BS--TO'' (17581--075),
one as ``SG'' (subgiant; 17581--113), one as ``TO'' (turn-off star; 22889--050),
and one as ``RG'' (red giant; 30325--028).

Table 2 shows the 177 VMP stars observed at La Silla with the DFOSC, $uvby$
photometry only.  Columns 1--6 are the same as in Table 1.  Column 7 
provides a few notes concerning the possible photometric variability (``V?'') of 
these VMP stars; below, three of these possible variables are 
classified ``HB'' (22881--039, 30339--046, and 31061--057) and so are candidates for 
VMP RR-Lyrae-type stars.  Two others are classified ``RHB--AGB'' (22952--015 and 
31083--069) and may be AGB semiregular or irregular variable stars, and one as 
``RG'' (22873--166).

\subsection{Comparisons with the HK-survey photometry}

For many of the stars in Tables 1 and 2, $UBV$ photometry has been obtained
as part of the HK survey of Beers and colleagues (see the references and 
Table 3 in the following section).
In order to check the quality of their data and also of our $uvby$--$\beta$
photometry, Fig.~1 shows the agreement between the two sets of $V$
magnitudes, and Fig.~2 the relation between the observed $(B$--$V)$ and our 
observed $(b$--$y)$.  Figure 1 shows the difference $V$(HK Survey)$-V$($uvby$)
as a function of $V$($uvby$) for the 419 stars in Tables 1 and 2 and in S96
for which $UBV$ photometry has been obtained.  
In this figure the CH stars indicated below in Col.~12 of Table 3 are 
graphed as filled circles; all others as open circles.  The 
overall distribution of points around the 0\fm00 line looks very satisfactory 
except for a few outliers. Five of the more extreme outliers have been 
marked with their names, and these all have differences in the observed $V$
magnitudes greater than 0\fm35.  Of these, 29512--013 and 30312--062 are 
from S96, both are classified below as ``HB'', and in S96,
29512--013 has been noted as ``variable''; probably both of these are
variable VMP RR-Lyrae-type stars.  As discussed below, 16551--118 and 
22180--036 may be binaries or variable AGB stars with anomalous abundances; 
both have $m_{\rm 1}$ values much too large to correspond to their 
[Fe/H]$_{\rm F}$ values of Table 7.  The star 31081--049 was observed at 
La Silla, has been classified as ``RG'', and is one of the reddest stars 
observed in this project.

   \begin{figure}
   \centering
   \includegraphics[width=9cm]{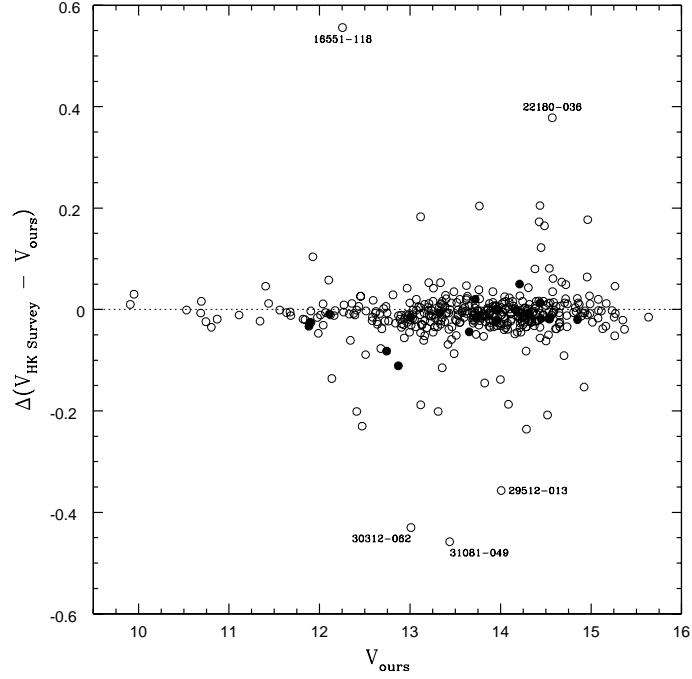}
      \caption{Comparison of the $V$ magnitudes: the difference 
               between the $V$ magnitudes of the HK survey minus 
               those of the present publication are plotted as a 
               function of the present $V$ values.  The dotted line
               shows the zero-difference level, and five of the more
               extreme outliers are marked with their identification
               numbers; these are discussed in the text.  Stars
               identified as CH stars in the HK survey (Col.~12
               of Table 3) are plotted as filled circles; all others
               as open circles.
              }
         \label{FigDeltaVmag}
   \end{figure}
%
   \begin{figure}
   \centering
   \includegraphics[width=9cm]{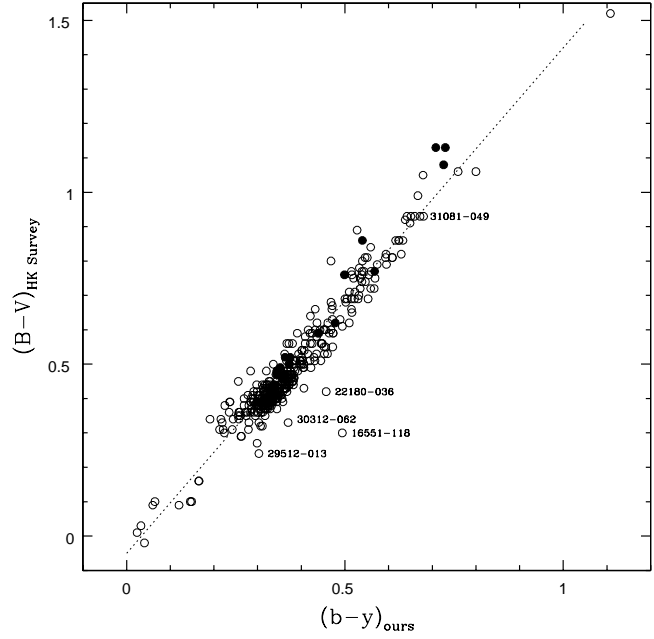}
      \caption{The $(B$--$V)$ colors observed for the HK survey 
               are graphed against
               the $(b$--$y)$ colors of the present publication.
               The dotted line has a slope of 1.47, as suggested by
               Budding (1993) for metal-free stars.  As in Fig. 1,
               the same five extreme outliers are labeled with their
               identification numbers, and the CH stars are plotted
               as filled circles.
              }
         \label{FigBVby}
   \end{figure}
%

Figure 2 shows the observed $(B$--$V)$ from the HK survey versus $(b$--$y)$
from our observations, for the 419 VMP stars which have been measured with 
the two systems.  Again the CH stars indicated in Table 3 have been plotted as 
filled circles, and all others as open circles.  The five outliers of the 
previous figure have again been labeled.  The dotted line has a slope of 
1.47, which is the approximate ratio between the observed $(B$--$V)$ and 
$(b$--$y)$ colors expected for metal-free stars, as suggested by the theoretical 
calculations of Budding (1993), and it can be seen that these data
do follow well this slope, as confirmed below in Fig.~4 and Eq.~1.  
Several of the CH stars are seen as outliers above this dotted line; this 
is not surprising since enhanced CN and CH absorptions decrease the flux 
in the $B$ band but do not affect $V$, $b$, and $y$.  Once again, 16551--118, 
22180--036, 29512--013, and 30312--062 are seen as outliers, probably due to 
the reasons mentioned elsewhere:  photometric variability, anomalous
chemical abundances, and/or a binary companion.

There are 77 HK survey stars in common between our (new) measurements of Str\"omgren
photometry from SPM and La Silla with the previously published work of
Anthony-Twarog et al.~(2000; AT).  The measurements of AT are not expected to be
as accurate as those reported in our present work, in part due to the fact that
their data was obtained with the CCDPHOT CCD-based detector system, which is
known not to provide an ideal match to the Str\"omgren system, and often they
only had single observations of each target.  Nevertheless, the agreement
between the two sets of data is quite respectable:

$$ <V_{\rm AT} - V_{\rm P}> = -0\fm019; \;\;\; \sigma (V_{\rm AT}-V_{\rm P}) = 0\fm057$$
$$ <(b-y)_{\rm AT} - (b-y)_{\rm P}> = +0\fm009; \;\;\; \sigma ((b-y)_{\rm AT}-(b-y)_{\rm P}) = 0\fm032$$
$$ <m_{\rm 1,AT} - m_{\rm 1,P}> = -0\fm011; \;\;\; \sigma (m_{\rm 1,AT}-m_{\rm 1,P}) = 0\fm040$$
$$ <c_{\rm 1,AT} - c_{\rm 1,P}> = +0\fm010; \;\;\; \sigma (c_{\rm 1,AT}-c_{\rm 1,P}) = 0\fm084$$

\noindent where the subscript P represents the present data, not discriminating
whether it came from SPM or La Silla.   If one uses robust and resistant
estimates of the central location and scale of the differences between the
measurements, e.g., as described by Beers, Flynn \& Gebhardt (1990), the
comparison is even more favorable:

$$ C\,(V_{\rm AT} - V_{\rm P}) = -0\fm015; \;\;\; S\,(V_{\rm AT}-V_{\rm P}) = 0\fm024$$
$$ C\,((b-y)_{\rm AT} - (b-y)_{\rm P}) = +0\fm008; \;\;\; S\,((b-y)_{\rm AT}-(b-y)_{\rm P}) = 0\fm026$$
$$ C\,(m_{\rm 1,AT} - m_{\rm 1,P}) = -0\fm009; \;\;\; S\,(m_{\rm 1,AT}-m_{\rm 1,P}) = 0\fm036$$
$$ C\,(c_{\rm 1,AT} - c_{\rm 1,P}) = +0\fm002; \;\;\; S\,(c_{\rm 1,AT}-c_{\rm 1,P}) = 0\fm051$$

\noindent Both the offsets in the mean values and the estimated rms variations are
consistent with expectations, considering the reported errors in both samples.

\section{Reddening and estimation of broadband colors}

Table 3 lists the positions of our program stars, both equatorial and Galactic,
along with broadband $V$ and $B-V$ photometry, where available. The sources for
this photometry include Doinidis \& Beers (1990), Doinidis \& Beers (1991),
Preston, Shectman, \& Beers
 (1991), Bonifacio, Monai, \& Beers (2000), and Beers
et al.~(2003, in preparation). In many cases, several sources have been
averaged. The typical accuracy of this photometry is on the order of $\sigma
(V)$ and $\sigma (B$--$V) \approx 0\fm01$--$0\fm02$.
  The stars in this 
and the following tables include those VMP stars from Tables 1 and 2 above, 
plus the VMP stars from S96.  In Table 4 are shown cross-identifications for
a number of the VMP stars; these are stars identified as VMP in more than
one of the overlapping fields from the HK survey.

Also listed in Col.~8 of Table 3 are the reddenings in the stellar 
directions obtained by
 interpolation in the maps of Schlegel, Finkbeiner, 
\& Davis (1998), which have
 superior spatial resolution and are thought to have 
a better-determined zero
 point than the Burstein \& Heiles (1982) maps. However, 
Arce \& Goodman (1999)
 caution that the Schlegel et al.~map may overestimate the 
reddening values when
 the color excess $E(B$--$V)_{\rm S}$ exceeds about 0\fm15. 
Our own independent
 tests suggest that this problem may extend to even lower 
color excesses, on the
 order of $E(B$--$V)_{\rm S} = 0\fm10$. Hence, we have 
adopted a slight
 revision of the Schlegel et al.~reddening estimates, according 
to the following:

\begin{equation}
\begin{array}{lclcl}
E(B-V)_{\rm A} & = & E(B-V)_{\rm S} &\;\;\;\;\; & E(B-V)_{\rm S} \le 0\fm10\\ [.25in]
E(B-V)_{\rm A} & = & 0\fm10 + 0.65 \times [E(B-V)_{\rm S} - 0\fm10] &\;\;\;\;\;&  E(B-V)_{\rm S} > 0\fm10
\end{array}
\end{equation}

\noindent where $E(B$--$V)_{\rm A}$ indicates the adopted reddening estimate.
We note that for $E(B$--$V)_{\rm S} \ge 0\fm15$ this approximately reproduces the
30\%--50\% reddening reduction recommended by Arce \& Goodman.  
The final 
adopted reddening, denoted as $E(B$--$V)_{\rm A}$, is given in Col.~9 of Table 3.
Column 10 lists the dereddened color $(B$--$V)_{\rm 0}$, for the stars where this
information is available.

These stars can also be dereddened using the intrinsic-color calibration of
Schuster \& Nissen (1989) when a value has been observed for H$\beta$, as for
most of the brighter VMP stars.  This calibration, plus a small offset correction
as noted by Nissen (1994), has been used to estimate interstellar reddenings
for 177 of the VMP stars being studied here.  In Fig.~3 a comparison of
$E(B$--$V)_{\rm A}$ with $E(b$--$y)$ for these 177 stars has been plotted.  
The dotted line shows the expected relation of $E(B$--$V) = 1.35E(b$--$y)$ (Crawford 
1975a), and indicates generally good agreement between the two dereddening methods, 
with no obvious systematic differences.  Table 5 lists the dereddened 
$uvby$ photometric values for all of our program stars:  Col.~1 the stellar 
identification number, Cols.~2--5 the values for $V_{\rm 0}$, $(b$--$y)_{\rm 0}$, 
$m_{\rm 0}$, and $c_{\rm 0}$, respectively, Cols.~6--7 the values of $E(b$--$y)$, 
from the intrinsic-color calibration when available, and 
$E(B$--$V)_{\rm A}$, as discussed above, respectively, and in 
Col.~8 the photometric classification to be discussed below.  Stars which
appear on more than one line have an asterisk following these classifications:
nine stars were observed at both SPM and La Silla, and, as mentioned above,
22955--032 was observed in two ways.  The dereddened photometry has been 
obtained by applying preferentially the $E(b$--$y)$ values from the 
intrinsic-color calibration of Schuster \& Nissen (1989), when H$\beta$ is 
available, or, if not, from $E(b$--$y)= E(B$--$V)_{\rm A}/1.35$.  Reddening 
corrections have been applied to the $uvby$ photometry only when 
$E(b$--$y) \geq 0\fm015$; values smaller than this are mostly not real but due 
to the photometric observational errors (see Nissen 1994).  For the other 
reddening corrections, these relations have been used:  $A_{\rm V} = 4.3E(b$--$y)$, 
$E(m_{\rm 1}) = -0.3E(b$--$y)$, and $E(c_{\rm 1}) = +0.2E(b$--$y)$ 
(Str\"omgren 1966; Crawford 1975a).

   \begin{figure}
   \centering
   \includegraphics[width=9cm]{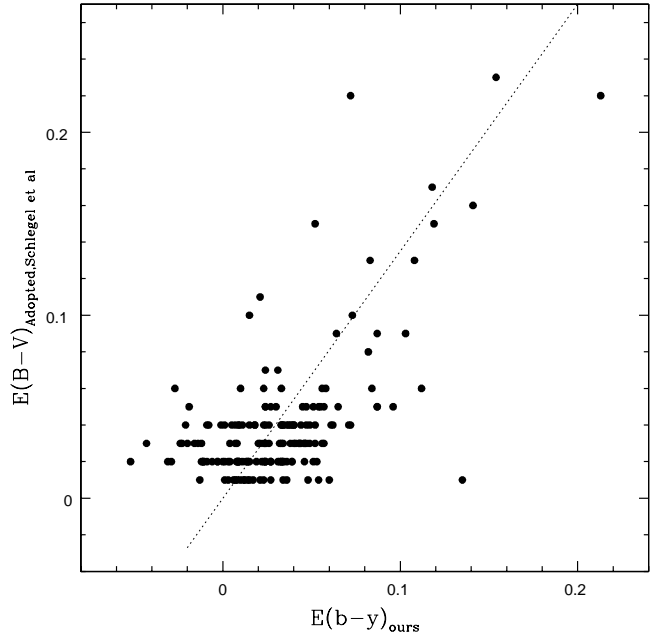}
      \caption{Comparison of the $E(B$--$V)_{\rm A}$ reddenings, as 
               discussed above, with the $E(b$--$y)$ values from the
               intrinsic-color calibration of Schuster \& Nissen (1989). 
               The dashed line has a slope of 1.35 (Crawford 1975a).
              }
         \label{FigEBVAEby}
   \end{figure}
%

Str\"omgren $(b$--$y)_{\rm 0}$ values are available for
 all of our target 
stars, and hence it is desirable to make use of this information
 to assist 
in the determination of the metallicity estimates.  Since the
 calibration of 
Beers et al.~(1999) is employed, an estimated
 $(B$--$V)_{\rm 0}$ color is 
first required.  Figure 4 shows a comparison of the dereddened broad- and 
intermediate-band
 photometry 
for the stars where both pieces of information are available.  The
 filled 
circles indicate stars which are likely outliers, as seen in Figs.~1 and 4,
with differences of more than 0\fm10 or 0\fm15, respectively, 
or were noted to have 
rather strong CH G-band indices, suggesting that they 
are carbon-enhanced (the CH stars).  The
 regression line, obtained using 
the stars which are not rejected for these
 reasons, is:  

$$(B-V)_{\rm 0} = 1.464 (\pm 0.022) \;(b-y)_{\rm 0} - 0\fm060 (\pm 0\fm008)$$

This slope is in very good agreement with that predicted theoretically by
Budding (1993) for metal-free stars, 1.47.  The expected errors in predictions 
of broadband color from application of this
 line are on the order of 0\fm05.  
This relation is used to obtain an
 estimate for the $(B$--$V)_{\rm 0}$ color, 
designated $BV_{\rm 0}$ and listed in Col.~11 of
 Table 3.  The final two 
columns of Table 3 indicate whether the star is noted
 as a carbon-enhanced 
star, or a photometric outlier as defined above, respectively.

   \begin{figure}
   \centering
   \includegraphics[width=9cm]{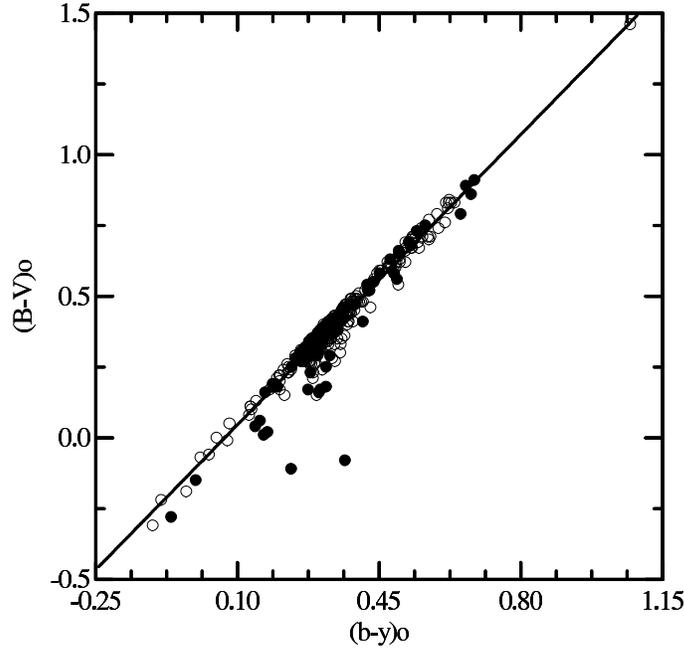}
      \caption{The $(B$--$V)_{\rm 0}$ colors are plotted against the 
               $(b$--$y)_{\rm 0}$ colors.  The diagonal solid line has 
               a slope of 1.464.  Probable outliers and the CH stars 
               are plotted as filled circles.
              }
         \label{FigBV0by0}
   \end{figure}
%

\section{Stellar abundances}

\subsection{Line indices}

Key line-strength indices for each of our stars have been measured using the
techniques, and bands, described in Beers et al.~(1999). These indices are
listed in Table 6. KP is the index that measures the strength of the CaII K
line, which serves as our primary metallicity indicator. HP2 and HG2 are indices
measuring the strengths of the Balmer lines H$\delta$ and H$\gamma$,
respectively. GP is an index that measures the strength of the CH G-band.

\subsection{Estimation of [Fe/H] values}

The stellar metallicities for our program objects are estimated with several
approaches. First, the estimated broadband color, $BV_{\rm 0}$ (in distinction
to a measured $(B-V)_{\rm 0}$) is used along with the CaII K line index, KP, to
obtain estimated metallicities for stars in the color range $0\fm3 \le BV_{\rm
0} \le 1\fm2$, based on the calibration of Beers et al.~(1999). This approach, a
multiple regression over the calibration space, has been demonstrated to provide
metallicities which are precise to about $\pm 0.2$ dex.  These values
are listed as [Fe/H]$_{\rm K1}$ in Col.~2 of Table 7.

As an alternative, abundance estimates have also been made based on an
Artificial Neural Network (ANN), using as inputs $BV_{\rm 0}$ and the base-ten
logarithm of the CaII K-line index, log(KP). This network was trained using the
same set of calibration stars as in Beers (1999), so it is not entirely
independent of the first method, but it does provide some information on errors
that might arise from the regression approach. The training process indicated
that the expected errors of prediction for metallicities derived from this
method should be on the order of 0.20--0.25 dex. This estimate is listed as
[Fe/H]$_{\rm A1}$ in Col.~3 of Table 7.

Similar estimates of abundances for program stars with available (\emph {measured})
broadband $(B$--$V)_{\rm 0}$ colors in the range $0\fm3 \le (B$--$V)_{\rm 0} \le
1\fm2$ are also obtained. The first approach, based on the Beers et al.~(1999)
calibration, and using $(B$--$V)_{\rm 0}$ and KP as inputs, yields the
metallicity estimates designated as [Fe/H]$_{\rm K2}$ in Col.~4 of Table 7.
The ANN estimate, based on log(KP) and $(B$--$V)_{\rm 0}$ is designated as
[Fe/H]$_{\rm A2}$ in Col.~5 of Table 7.

Inspection of Table 7 shows that the four estimates of metallicity are often,
though not always, in rather good agreement with one another. The most
discrepant cases arise for stars where the estimated $BV_{\rm 0}$ and measured
$(B$--$V)_{\rm 0}$ colors disagree. Final estimates of metallicity are assigned,
in general, based on a straight average of the individual abundances, and are
designated as [Fe/H]$_{\rm F}$ in Col.~6 of Table 7. In a few cases, 
preference was given to one or more of the individual estimates; this in 
particular applies to the cooler, more metal-rich stars.  The Beers et al. 
(1999) procedure applies an explicit correction for saturation effects in the 
KP index, which the ANN procedure does not.

\subsection{Carbon-enhanced stars}

There are a number of stars in our sample that clearly exhibit enhanced carbon
abundances, as demonstrated from the strengths of their CH G-band indices, GP.
Such stars have been noted in a number of recent studies (e.g., Norris, Ryan,\&
Beers 1997; Za\u{c}s, Nissen, \& Schuster 1998; Rossi, Beers, \& Sneden 1999) to
occur with a larger frequency amongst stars of very low metallicity, as compared
to stars of intermediate and solar abundance. These stars also provide important
probes of early stellar evolution at low metallicity (e.g., Fujimoto, Ikeda, \& Iben
2000; Schlattl et al. 2002), as well as operation of the s-process in the early
Galaxy (e.g., Aoki et al.~2002a; Aoki et al.~2002c). In Figure 5, the GP index
is plotted versus the KP index for our program stars. The stars that
clearly stand out from the rest of the sample are marked with filled circles,
and are likely carbon-enhanced stars (these stars are also noted in Table 3).
Note that a number of these stars have already had detailed studies of their
abundances, and in some cases, orbital properties, in the published literature. 

   \begin{figure}
   \centering
   \includegraphics[width=9cm]{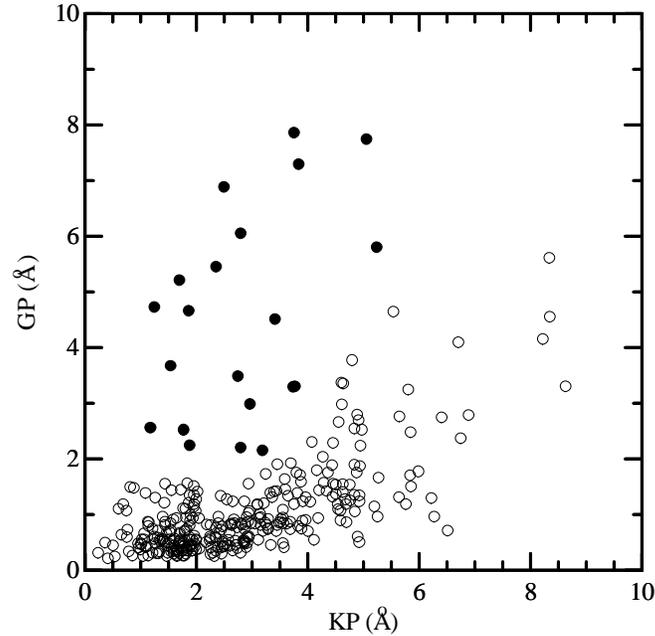}
      \caption{Comparison of the GP index with the KP. 
               Probable CH stars are plotted as filled circles.
              }
         \label{FigGPKP}
   \end{figure}
%

\subsection{Other possible abundance anomalies}

In Table 5 are seen a number of stars with $(b$--$y)_{\rm 0} \la 0\fm45$ and 
with $m_{\rm 0} \ga 0\fm17$, much larger than would be expected for VMP 
stars with [Fe/H] $\la -1.5$.  (For example, see Fig.~4 from Schuster \&
Nissen 1989).  Examples of such stars are 16548--009, 16551--118, 
16552--086, 17572--057, 17586--014, 22176--018, 22180--036, 31081--003, 
and 31083--069.  Most of these stars (except 16548--009 and 17572--057) 
are also noted as outliers in Table 3 with differences greater 
than 0\fm10 and/or 0\fm15 in Figs.~1 and 4, respectively; 16551--118 and 
22180--036 are also two of the more extreme outliers labeled in Figs.~1 
and 2.  These stars may be lower-temperature analogues to the eight stars
plotted in Fig.~5 and discussed in Sect.~3.1 of S96, those with larger than 
expected $[m_{\rm 1}]$ values, explained as having ``... some 
anomaly, such as an unusual chemical abundance ratio or a binary companion.''  
Also in S96 it was noted that BPSII had identified four of these previous 
stars as having ``... unusually strong G bands and CN features,'' but in Table 3 
none of the above mentioned stars have been identified as CH stars.  All of the 
present cases (except 31083--069, which has been classified RHB--AGB) have been 
classified subgiants (SG) in Fig.~6, and also several have the added dimension of 
probable photometric variability (such as 16551--118 and 22180--036).  In the 
$c_{\rm 0},(b$--$y)_{\rm 0}$ classification diagram of Fig.~6 the SGs and 
RHB--AGBs span nearly the same range in $(b$--$y)_{\rm 0}$, and it has been noted 
in the literature that the AGB stars sometimes exhibit photometric variability, 
being Mira, semiregular-, or irregular-type variables (Gautschy \& Hideyuki 1996) 
and frequently anomalous chemical abundances (Busso et al.~1999).  It is suggested 
here that many of these above-mentioned SG stars may in fact be mis-classified 
(variable) AGB stars with unusual chemical abundance ratios.   It is known, for 
example, that nitrogen variations can shift the $c_{\rm 0}$ index via the effect 
of the NH band at $3360${\AA}, and the most convincing demonstration of 
this has been given by Grundahl et al.~(2002), who studied red giants in NGC6752.  
In addition, other abundances may also affect the $c_{\rm 0}$ index.  For example, 
Za\u{c}s et al.~(1998) discuss the probable effects of CH upon the $v$--band of 
Str\"omgren photometry, and Grundahl et al.~(2000b) have shown that scatter in 
$c_{\rm 1}$ is seen in globular clusters down to at least the base of the 
red-giant branch and that this scatter is correlated with the CN strength.

Another group of stars which stand out quite clearly using the $uvby$ photometry
together with the indices of the HK survey are those 10 stars of Table 5 with 
the classification ``BS (Am)''.  These are discussed in more detail below in 
Sect.~5.2.  These are stars which appear to have an underabundance 
of Ca which does not correspond to the abundance of Fe.  For example, Wilhelm 
et al.~(1999a, 1999b) work with two Ca II K-line estimators to derive [Fe/H], 
and also two metallic-line regions which include Fe and Mg lines.  For the 
Am stars these different indicators can give widely different [Fe/H] values, 
such as for example the stars 22871-0111, 22956-0055, and 30321-0076 from Table 2A 
of Wilhelm et al.~(1999b), which have [Fe/H] $\la -2.0$ from the K-line estimators 
and [Fe/H] $\approx 0.00$ from the metallic-line regions.  The ``BS (Am)'' stars 
of this paper show exactly these characteristics, as discussed below.

\section{Photometric Classifications}
\subsection{The $c_{\rm 0}$, $(b$--$y)_{\rm 0}$ diagram}

In S96 the $[c_{\rm 1}],[m_{\rm 1}]$; $c_{\rm 0},(b$--$y)_{\rm 0}$; and 
$[c_{\rm 1}],\beta$ diagrams were used to derive and analyze 
the photometric classifications
of the VMP stars.  However, for the present work fewer stars have H$\beta$
values, making $[c_{\rm 1}]$,$\beta$ less useful.  Also, here the range in
metallicities is wider than for S96, making the classifications from the 
$[c_{\rm 1}]$,$[m_{\rm 1}]$ diagram more difficult and less definitive;
$[m_{\rm 1}]$ is sensitive to both metallicity and temperature and so is less 
adequate as the second classification parameter.  So, for the present work the 
photometric classifications have been obtained using mainly the 
$c_{\rm 0}$,$(b$--$y)_{\rm 0}$ diagram with the $[c_{\rm 1}]$,$[m_{\rm 1}]$ 
diagram used only for some checking.  All of the present VMP stars have been 
observed in $c_{\rm 1}$ and $(b$--$y)$, all have been dereddened using the 
methods discussed above, and the metallicity effects upon $c_{\rm 0}$ and 
$(b$--$y)_{\rm 0}$ are small, especially for the A- and F-type stars.

Our photometric classification scheme for the VMP stars is based mainly upon
three sources:  the separation of candidate VMP subgiant stars by Pilachowski
et al.~(1993, Fig.~1), the study of halo red giant, AGB, and horizontal-branch
stars by Anthony-Twarog \& Twarog (1994, Fig.~9), and a large amount of $uvby$
photometry for globular clusters (hereafter GCs) provided by Frank Grundahl (2000).  
The final classification diagram is shown in Fig.~6.  The first iteration of this 
diagram was derived using the first two sources mentioned above, Pilachowski 
et al.~(1993), and Anthony-Twarog \& Twarog (1994), and then several refinements 
and extensions were provided by the $uvby$ data from Grundahl (2000).  For example,
the dividing line between the turnoff (TO) and subgiant (SG) stars was first
taken from Pilachowski et al.~but then modified slightly using the $uvby$ data 
from Grundahl for the GC M92.  The data for M92 was also used to refine and 
extend the classification area for the red-horizontal-branch-asymptotic-giant-branch 
(RHB--AGB) stars and for the blue horizontal branch (BHB).
Data for M2 and NGC1851 were used to better define the area of the horizontal
branch (HB), and NGC6752, M79, and M13 for the subluminous-blue-horizontal-branch 
(SL--BHB).  The other classification categories are:  main sequence (MS),
blue straggler (BS), red giant (RG), and subluminous (SL).

Photometric classifications for the VMP stars, derived from this Fig.~6, are
given in the last column of Table 5 along with the $c_{\rm 0}$ and 
$(b$--$y)_{\rm 0}$ values used, Cols.~5 and 3, respectively.  These 
classifications are also repeated in Table 9.  Following these classifications, 
in parentheses, are given indications of abundance anomalies, such as ``Am" 
from Sects.~4.4 and 5.2, the ``CH" stars of Sect.~4.3 and Table 3, and the 
``CNO" and ``CN" stars, 22949--037 and 29498--043, respectively, from Sect.~6.2.  
One should expect that the classifications and distances of these anomalous 
stars are less reliable.  In Table 5 excellent agreement is seen for stars 
observed at both SPM and La Silla, those with asterisks at the end (except 
22955--032); the photometric classifications are identical in all nine cases.  
We emphasize that these classifications are photometric and correspond most 
closely to those classifications of metal-poor stars from the color-magnitude 
diagrams of GCs, rather than to any spectroscopic classification.

The star 22955--032 provides a good example of the possible classification
errors produced by the photometric contamination of a nearby star.  As 
mentioned above, this star was observed in two ways, with a nearby,
fainter, apparently redder, star both included and excluded from the
photometer's diaphragm.  From the one less-contaminated observation
22955--032 is classified as TO, for the two contaminated observations as MS,
and for the combined, three observations as SG.

   \begin{figure}
   \centering
   \includegraphics[width=12cm]{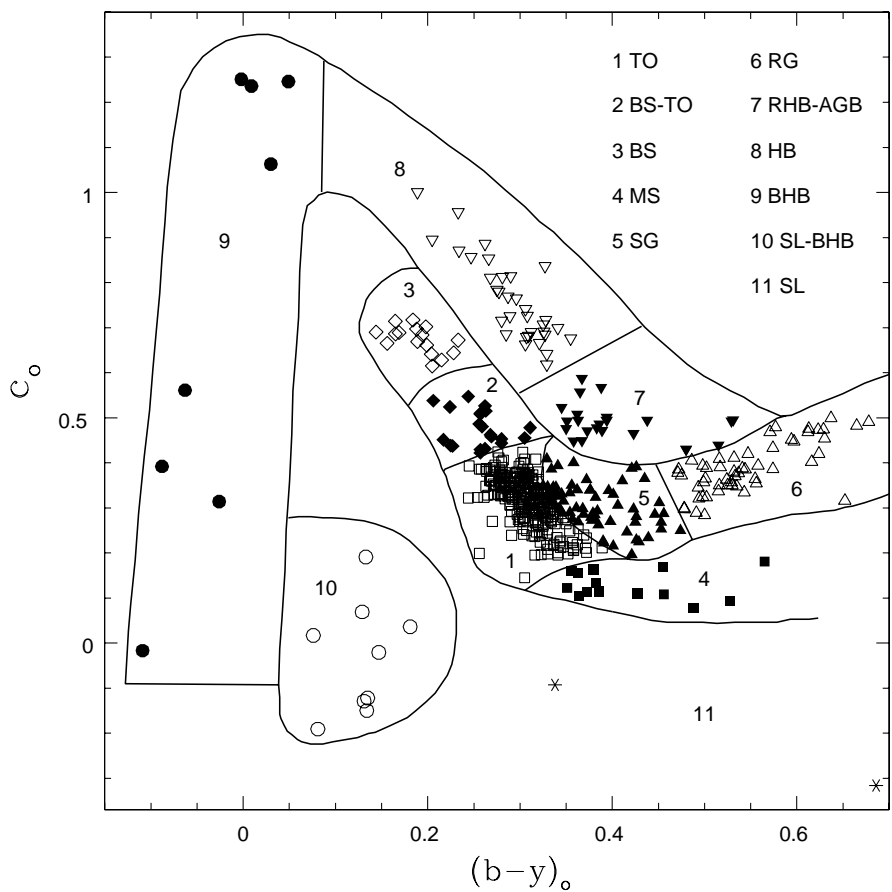}
      \caption{The $c_{\rm 0}$,$(b$--$y)_{\rm 0}$ diagram for the VMP 
               stars of this paper with the classification areas 
               indicated.  Area 1 corresponds to turnoff stars (TO); 
               area 2, the blue-straggler--turnoff transition (BS--TO);
               area 3, blue stragglers (BS); area 4, the main sequence
               (MS); area 5, subgiant stars (SG); area 6, red giants
               (RG); area 7, the 
               red-horizontal-branch--asymptotic-giant-branch transition 
               (RHB--AGB); area 8, the
               horizontal branch (HB); area 9, the blue horizontal
               branch (BHB); area 10, the 
               subluminous--blue-horizontal-branch transition (SL--BHB); 
               and area 11, the subluminous (SL)
               stars.  These are photometric classifications
               corresponding most closely to the stellar categories 
               from GC color-magnitude diagrams.
              }
         \label{Figc0by0}
   \end{figure}
%

\subsection{The $[c_{\rm 1}]$, $[m_{\rm 1}]$ diagram}

The $[c_{\rm 1}]$, $[m_{\rm 1}]$ diagram has also been plotted (not shown)
for all of the VMP stars from Tables 1 and 2, plus those from S96, where
$[c_{\rm 1}] = c_{\rm 1} - 0.20(b$--$y)$ and $[m_{\rm 1}] = m_{\rm 1} + 0.30(b$--$y)$
are reddening-free indices according to the work of Str\"omgren (1966) and
of Crawford (1975a).  Due to the sensitivity of $[m_{\rm 1}]$ to metallicity and 
due to the considerable range in our sample from [Fe/H] $\approx -0.9$ to 
$\approx -3.8$, this diagram is not easily used to classify the VMP stars, 
but many of the features of Fig.~6 can be traced, such as the turnoff, subgiant, 
blue straggler, horizontal-branch, and subluminous stars.  

What does stand out in this figure is a compact group of ten stars with 
$<[c_{\rm 1}]> = 0\fm650 \pm 0\fm021$ and $<[m_{\rm 1}]> = 0\fm286 \pm 0\fm011$ 
falling to the red in $[m_{\rm 1}]$ of the horizontal-branch and blue-straggler 
stars.  The compactness and separation of this group is obvious, and the shift 
to more positive $[m_{\rm 1}]$ values would indicate higher metallicities. 
In Fig.~6 these ten have all been classified as blue stragglers (BS), and in 
Table 8 their photometric and physical properties are summarized, taken mostly 
from the above tables.  These stars have interstellar reddenings from 0\fm03 to 
0\fm39, Galactic longitudes from $68.8\degr$ to $244.0\degr$, latitudes from 
$-28.1\degr$ to $+39.1\degr$, and $(B$--$V)_{\rm 0}$ values which place them 
near the blue limit of the HK survey.  All are indicated as outliers in Table 3 
mostly due to their anomalous positions in Fig.~4.  Four of the stars, which 
have had their metallicities measured, have [Fe/H]$_{\rm F} \approx -2.0$, but 
the $m_{\rm 1}$, $[m_{\rm 1}]$, and $(U-B)_{\rm 0}$ values of these ten objects 
would all indicate much higher, nearly solar, metallicities.  The other six do 
not have [Fe/H]$_{\rm F}$ values, being bluer than 0\fm30 in $BV_{\rm 0}$ and 
$(B$--$V)_{\rm 0}$, when available.

These ten stars fit the definition of an ``Am'' star, as mentioned above in Sect.~4.4,
and as developed in Wilhelm et al.~(1999a, 1999b).  The KP index of these stars
would indicate low metallicities, [Fe/H] $\la -1.50$, while other metal-sensitive
indices from the $uvby$ and $UBV$ photometries would indicate nearly solar values.
In Table 5 a note ``(Am)'' has been attached to the ``BS'' classification of these
stars.

\section{Distance Estimates}
\subsection{Distances from $UBV$ photometry}

In order to obtain photometric estimates of the stellar distances, the
photometric classifications, based on the Str\"omgren indices listed in Table 5,
have first been adopted.  As noted in Sect.~5.1, those stars with indications
of abundance anomalies have less reliable classifications and consequently
less reliable photometric distances.  The $UBV$-based distances have been 
derived only from \emph{measured} broadband $V$ and $B-V$ photometry, where 
it exists.  The broadband de-reddened colors, $(B-V)_0$, and the final
averaged metallicities, [Fe/H]$_{\rm F}$, where available, are then used to
enter the procedures for obtaining estimates of the absolute magnitudes, M$_V$.
For stars classified as BHB, the relationship between absolute magnitude and 
metallicity adopted by Clementini et al.~(1995) has been used:

$$ M_{\rm V} {\rm (BHB)} = 0\fm68 + 0.19\;({\rm [Fe/H]} + 1.5).$$

For stars with other classifications, a discrepancy has been noted between the
absolute magnitudes obtained from the Revised Yale Isochrones employed by Beers
et al.~(1999) and those obtained based on calibrations of the Str\"omgren
photometry (described below). This discrepancy was most severe for the stars
classified as RG and AGB, in the sense that these stars were assigned absolute
magnitudes that were too bright. Hence, we decided to adopt the absolute
magnitudes as assigned by Beers et al.~(2000; their Table 2), based on
empirical calibrations of globular and open clusters, for the stars classified
as MS, TO, SG, RG, and AGB.  The adopted absolute magnitudes, and corresponding
distance estimates are listed in Table 9.  For the TO stars, absolute magnitudes 
and distances have been listed under the assumptions that the stars are either
main-sequence dwarfs or subgiants, and these absolute magnitudes and
distances derived under these assumptions are listed in Table 9
as $M_{\rm V1}$, $D_{\rm 1}$, and $M_{\rm V2}$, $D_{\rm 2}$, respectively.  
The great majority of the TO stars are expected to be dwarfs, and hence the 
primary estimates for these stars are $M_{\rm V1}$ and $D_{\rm 1}$.  For stars 
other than TO, the estimated
absolute magnitudes and distances are listed in the columns labeled $M_{\rm V1}$
and $D_{\rm 1}$, respectively.

\subsection{Distances from $uvby$--$\beta$ photometry}

Stellar distances have also been obtained from the $uvby$ photometry using a
variety of methods and calibration procedures, depending on the photometric
classifications given above.  These $uvby$ distances are given in the last 
column of Table 9.  For example, for the TO, MS, and most SG and BS--TO stars 
the $M_{\rm V}$ and photometric distances are derived from an empirical 
calibration based upon Hipparcos data (ESA 1997).  This calibration will be 
documented in greater detail elsewhere; here its characteristics are outlined 
briefly.  The calibration equation is based upon 512 stars from the Hipparcos 
data base with parallax errors of 10\% or less.  The Lutz-Kelker corrections 
to $M_{\rm V}$ for these stars are less than about 0\fm12.  This sample has been 
cleaned of binaries using other data bases and also by an iterative procedure 
whereby stars with residuals $\ga 0\fm7$ in the calibration comparison have been 
removed.  The calibration equation is a polynomial in $(b$--$y)$, $c_{\rm 0}$, 
and $m_{\rm 0}$ and higher-order terms to fourth order.  As for the 
calibrations of Schuster \& Nissen (1989), the solution has been iterated until 
all terms have T-ratios with absolute values greater than 3.  That is, all 
coefficients are at least three times their estimated errors according to the 
IDL REGRESS routine (the returned errors of the coefficients are standard 
deviations), and with $\sim 500$ degrees of freedom, all coefficients are non-zero 
at a significant level, greater than 0.995.  The 512 calibration stars have the 
following photometric ranges:  
$0\fm038 \leq m_{\rm 0} \leq 0\fm593$, $0\fm279 \leq (b$--$y) \leq 0\fm600$, 
$0\fm102 \leq c_{\rm 0} \leq 0\fm474$, and $0\fm991 \leq M_{\rm V} \leq 8\fm029$.  
(The actual region in the $M_{\rm V},(b$--$y)$ diagram over which this 
calibration is well-defined is a somewhat irregular polygon, and not a rectangle.)  
As mentioned in S96, for the VMP stars many of the photometric calibrations are 
not entirely adequate since few good calibration stars with [Fe/H] $< -2.0$ exist.  
This caveat also applies here, but this Hipparcos-based photometric calibration 
seems to work quite well for our VMP stars as suggested in Figs.~7 and 8.

In Fig.~7 are plotted $M_{\rm V}$ values calculated using this 
Hipparcos-based, empirical calibration against $M_{\rm V}$ derived directly 
from the Hipparcos parallaxes for nine of our calibration stars, those with the 
lowest metallicities, [Fe/H] $< -1.50$, according to the photometric [Fe/H] 
calibration of Schuster \& Nissen (1989).  These nine stars have [Fe/H] 
values in the range $-2.39 \leq$ [Fe/H] $\leq -1.57$.  The agreement seen 
in Fig.~7 is quite satisfactory, but as in the above caveat, these calibration 
stars do not extend to the lowest [Fe/H] values of many of the VMP stars.
In Fig.~8 the distances ($D_{\rm 1}$) and $M_{\rm V1}$ values from the 
HK survey, derived as described above using $UBV$ photometry, are compared 
to our distances and $M_{\rm V}$ values from our Hipparcos-based calibration.  
The comparison is shown for distances out to 2 kpc only, where our Hipparcos 
calibration dominates, and only TO, MS, SG and BS--TO stars have been plotted.  
Again the agreement seems quite good, with no indications for systematic 
problems with our Hipparcos-based, empirical calibration for $M_{\rm V}$.  
In the upper panel of Fig.~8 the mean locus of the TO stars is seen to be 
nearly horizontal and tilted with respect to the one-to-one dashed line.  
This is to be expected, since the $M_{\rm V1}$ values have been derived 
assuming that the VMP TO stars are all main-sequence dwarf stars while the 
Hipparcos-based calibration for $uvby$ photometry takes into account the 
evolution of these TO stars up to the subgiant branch.

%
   \begin{figure}
   \centering
   \includegraphics[width=9cm]{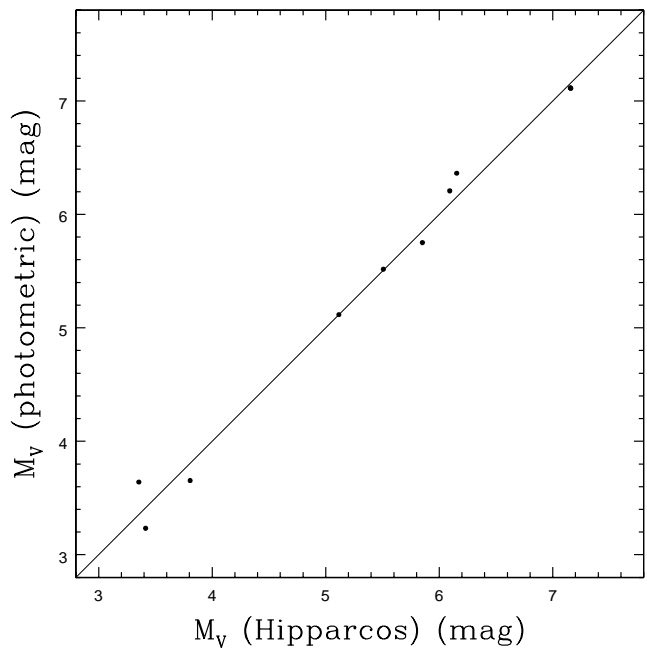}
      \caption{$M_{\rm V}$ values are compared for
               nine of our calibration stars with
               [Fe/H] $< -1.50$, according to the [Fe/H]
               calibration of Schuster \& Nissen (1989). 
               $M_{\rm V}$ calculated using $uvby$
               photometry plus our empirical, Hipparcos-based, 
               photometric calibration are plotted 
               against $M_{\rm V}$ values taken directly 
               from the Hipparcos parallaxes.  The solid, 
               diagonal line shows a one-to-one relation.
              }
         \label{FigMvsVMP}
   \end{figure}
%

   \begin{figure}
   \centering
   \includegraphics[width=9cm]{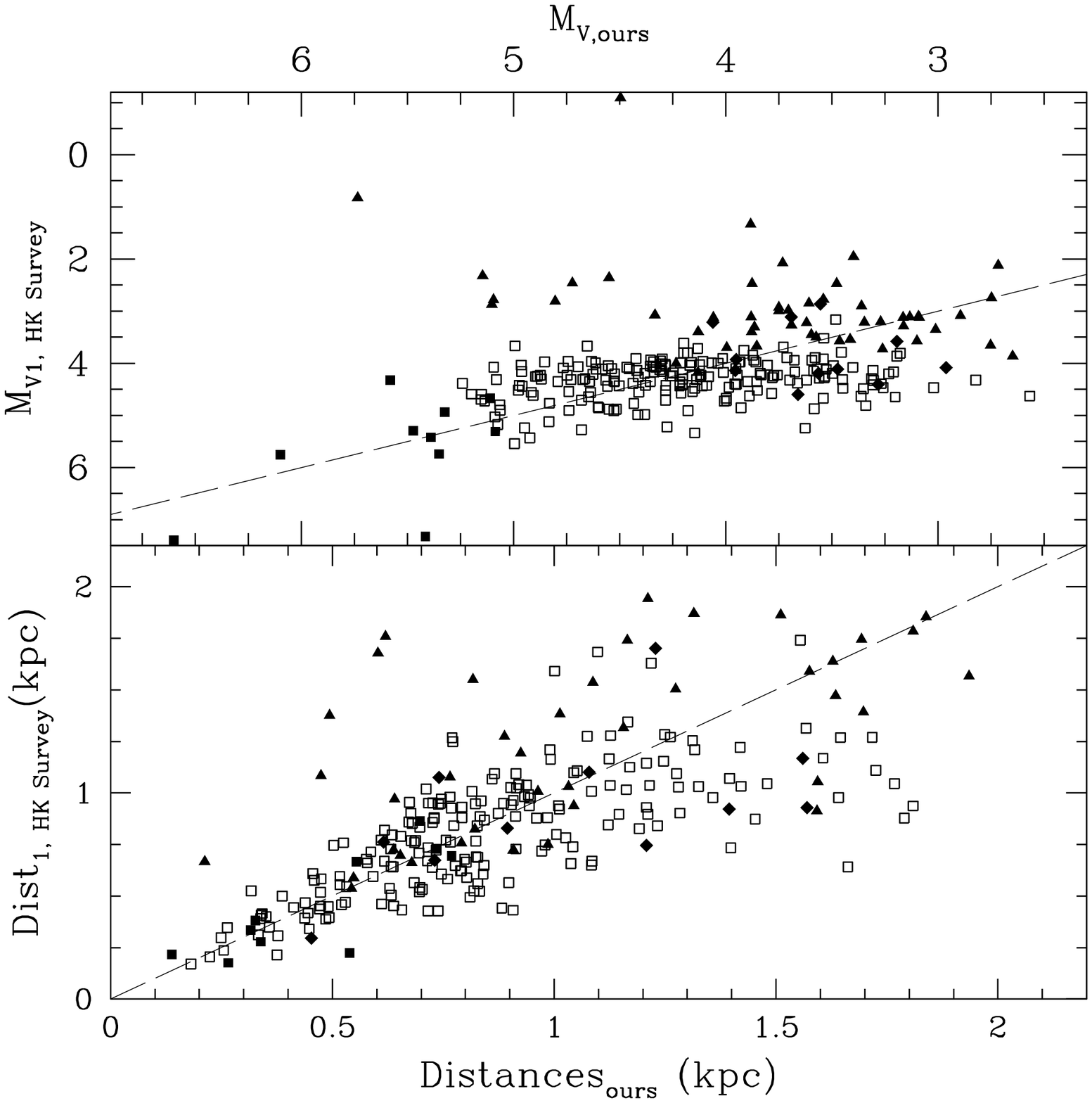}
      \caption{Comparison of distances ($D_{\rm 1}$, lower panel) 
               and $M_{\rm V1}$ (upper panel) from the 
               HK survey, derived from $UBV$ photometry as discussed 
               above, with those derived from $uvby$ photometry and
               our empirical, Hipparcos-based photometric 
               calibration for $M_{\rm V}$.  The dashed lines 
               show one-to-one relations.  The open squares 
               indicate TO stars, filled diamonds BS--TO, filled 
               squares MS, and filled triangles SG.  The comparison
               is shown for distances less than 2 kpc only, where 
               our Hipparcos-based photometric calibration dominates. 
              }
         \label{FigDist2kpc}
   \end{figure}
%

Our Hipparcos-based calibration can be applied over the ranges in $uvby$
photometry and $M_{\rm V}$ mentioned above.  Outside these ranges other
methods and calibrations have been used.  For the RGs and a few of the
brighter SGs, the stars have been fit to the color-magnitude diagram 
of M92, $(M_{\rm V})_{\rm 0}$  versus $(b$--$y)_{\rm 0}$, from Grundahl 
et al.~(2000a) using their distance modulus of 14\fm62 and $E(b$--$y) = 
0\fm016$, and also to the color-magnitude diagram of Grundahl et 
al.~(1998) for M13 with their distance modulus of 14\fm38 and 
$E(b$--$y) = 0\fm011$.  Then, assuming
[Fe/H] $\approx -2.3$ for M92 and [Fe/H] $\approx -1.6$ for M13, the
$M_{\rm V}$ value corresponding to the VMP star's [Fe/H] has been 
interpolated or extrapolated.  As a check, the models of Bergbusch \&
VandenBerg (2001) plus the color-temperature relations and isochrones of
Clem et al.~(2003) have been used to derive differential relations between
$\Delta M_{\rm V}$ and $\Delta$[Fe/H] for the RG stars, $\Delta M_{\rm V}
\approx 0.845\Delta$[Fe/H], and for the brighter SG stars, $\Delta M_{\rm V}
\approx 0.57\Delta$[Fe/H].  These differential relations are used together 
with the $M_{\rm V}$ measured from the color-magnitude diagram of M92 
(Grundahl et al.~2000a),
mentioned above, to again estimate $M_{\rm V}$ for the VMP RGs and 
brighter SGs.  For [Fe/H] values less than about $-2.3$, these differential 
relations must also be extrapolated.  For a very large majority of cases, 
these two methods gave $M_{\rm V}$ values which agree to within 
$0\fm10$--$0\fm15$.   The latter method has been used for most of the 
$M_{\rm V}$ adopted.

For the HB stars of our sample, the color-magnitude diagram of M92 from 
Grundahl et al.~(2000a), as provided by Grundahl (2000), has again been used, 
plus the relation $\Delta M_{\rm V} = 0.2\Delta$[Fe/H], from Harris (1994), 
as quoted by Kravtsov et al.~(1997), to correct from the [Fe/H] of M92 to that 
of the VMP HB star.  For the RHB--AGB and BHB stars, $M_{\rm V}$ has 
been read from the color-magnitude diagrams of M92 and M13, as referenced 
above, and then interpolated or extrapolated to the [Fe/H] of the VMP star.

For the BS and a few of the brighter BS--TO stars, two processes have been
employed depending upon [Fe/H].  For the metal-rich BS (Am) stars of Table 8,
the A- and F-star calibrations of Crawford (1975b, 1979) have been used
to derive $M_{\rm V}$.  Metal-poor BS have been compared 
to the blue stragglers of M3 (Rey et al.~2001) and of M13 (Grundahl et al.
1998) assuming their distance moduli of 14\fm93 and 14\fm38, respectively.  
These two clusters both have [Fe/H] $\approx -1.6$, and so again the the models 
of Bergbusch \& VandenBerg (2001) plus the isochrones of Clem et al.~(2003) 
were used to provide corrections to $M_{\rm V}$ as a function of [Fe/H] for 
stars near the main sequence.

$M_{\rm V}$ for the SL--BHB stars has been derived by assuming that 
they are
similar to the hot B subdwarfs studied by Villeneuve et al.~(1995) with $uvby$
photometry by Wesemael et al.~(1992), and also are like stars observed near the
lower end of the BHB in M13.  Then a comparison was made between the 
$(M_{\rm V})_{\rm 0},(b$--$y)_{\rm 0}$ and $c_{\rm 0},(b$--$y)_{\rm 0}$ 
diagrams for M13 using the $uvby$ data provided by Grundahl (2000). 
By analogy $M_{\rm V}$ of such VMP SL--BHB stars was deduced from 
a comparison with their $c_{\rm 0},(b$--$y)_{\rm 0}$ diagram.
There are only three SL stars in our sample, 17569--011, 22169--002, and
22948--027, the latter two are seen to be CH stars in Tables 3 and 5, and so
their actual nature is dubious.  We have assumed that they are white dwarfs, 
have taken their $(B$--$V)_{\rm 0}$ from the HK survey, and then derived 
$M_{\rm V}$ from Hansen \& Kawaler (1994) and Weidemann (1968).

In Fig.~9 are compared the HK-survey $D_{\rm 1}$ and $M_{\rm V1}$ from 
the $UBV$ photometry, as documented above, with the distances and 
$M_{\rm V}$ from $uvby$ photometry and the several
methods described above, over the full range of
application:  $\approx 0$--$16$ kpc.  In general the agreement
is quite good, considering the extrapolations necessary to calibrate
and derive distances for the more metal-poor VMP stars.  Some systematic
differences are noted for some of the groups, but these are within the
reasonable uncertainties of the calibration processes.  For example,
our HB distances are 5--10\% larger than those of the HK survey, and
our RG distances about 10\% larger.  The more discrepant stars in this
figure are all RG, SG, and RHB--AGB stars and may indicate photometric
variability, binary companions, and/or anomalous chemical compositions which 
affect the two photometries differently, as discussed above in Sect.~4.4.  
For example, the discrepant RG stars 22949-037 and 29498-043 are 
explained in the following paragraph.

   \begin{figure}
   \centering
   \includegraphics[width=9cm]{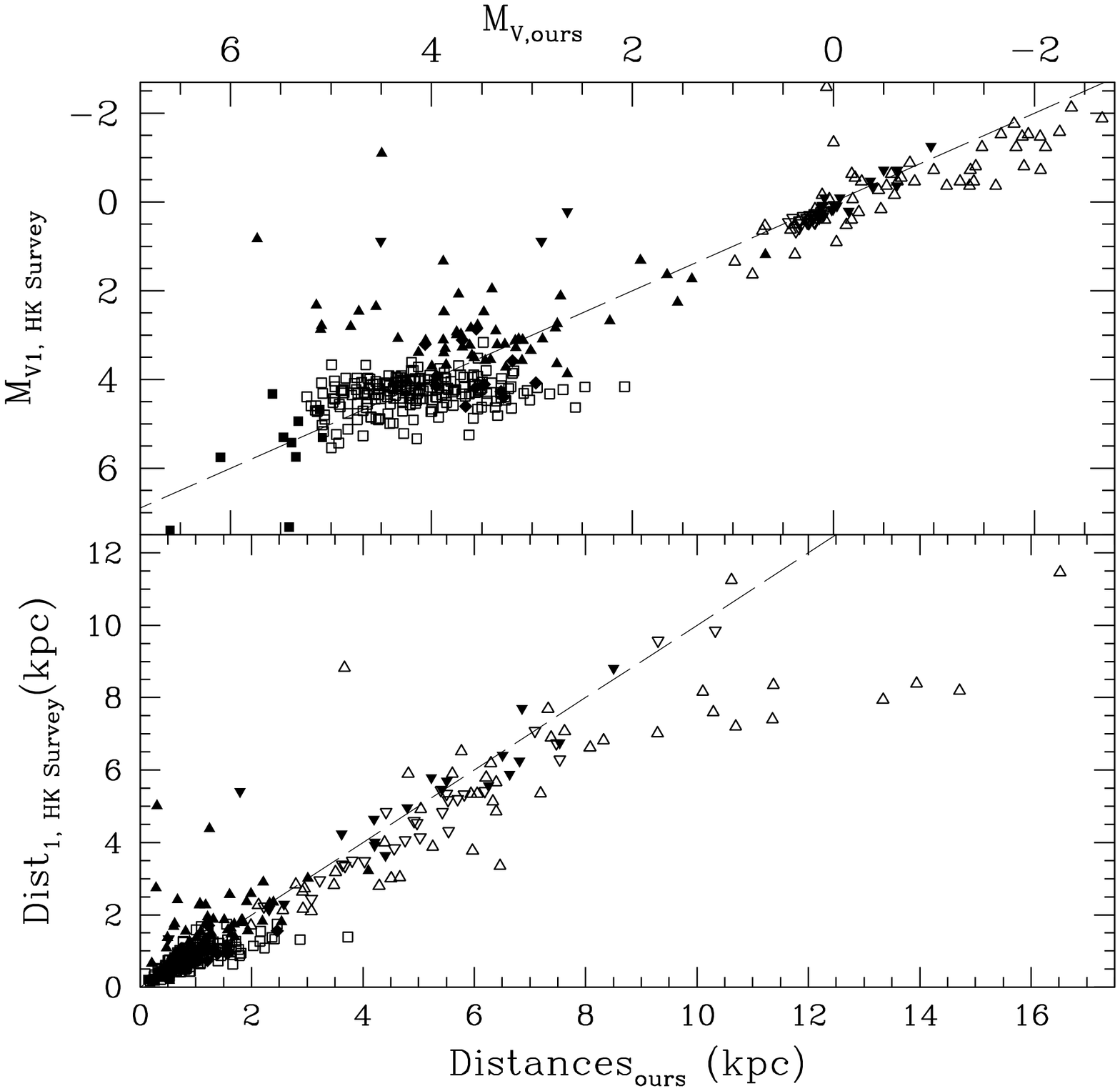}
      \caption{The same as Fig.~8 but for the full range of distances
               and absolute magnitudes.  The dashed lines show 
               one-to-one relations.  
               The open squares indicate TO stars, filled diamonds BS--TO, 
               open diamonds BS, filled squares MS, filled triangles 
               SG, open triangles RG, filled inverted triangles RHB--AGB,
               open inverted triangles HB, filled circles BHB, open
               circles SL--BHB, and asterisks the SL stars.  The 
               comparison is shown for all VMP stars of this paper 
               and of S96 for which distances have been derived from 
               the two photometric systems. 
              }
         \label{FigDistAll}
   \end{figure}
%

During analyses and comparisons which followed those of this paper, a few 
possibly discrepant stars and 
their distances have come to light.  For example, the distance of the RG star
22949--037 has been estimated at $\approx 14.7$ kpc from the $uvby$ photometry.  
This distance leads to very extreme Galactic velocities for this star, 
(U',V',W') $\approx$ ($-622$, $-1420$, $-849$), probably implying that it is not 
bound to the Galaxy (see Fig.~1 of Garcia-Cole et al.~1999). A more likely 
explanation is that the $c_{\rm 0}$ index is distorted by large CNO
overabundances (Depagne et al.~2002), leading to an unrealistic distance.  
(22949--037 has not been labeled as a CH star in Tables 3 and 5 since its 
medium-resolution spectrum of the HK survey did not extend far enough into 
the red to capture the G band.)  The VMP RG star 29498--043 has similar 
overabundances of C and N (Aoki et al.~2002a, 2002b), and so its
$uvby$ distance in Table 9 is probably similarly in error.  The stars 30339--041
and 31083--069 have both been classified as RHB--AGB stars, but their $uvby$
photometry falls within the limits of our Hipparcos-based calibration giving
$M_{\rm V}$ more like SG stars.  The explanation here may be anomalous chemical 
abundances for AGB-like stars, binary contamination, or photometric variability, 
especially the latter for 31083--069 for which the SPM and La Silla $uvby$ 
observations do not agree well.  The stars 29515--060 (classified MS) and 
31081--003 (classified SG) have widely different $M_{\rm V}$ values from 
different methods which should be applicable, such as the Hipparcos-based 
calibration and the comparison with GCs.  The former shows a 
difference $\Delta M_{\rm V} \approx 1\fm06$, and the latter $\approx 3\fm6$, and
these differences may again imply anomalous chemical abundances or a 
binary companion for these two stars.  As discussed above the $uvby$ data for 
22955--032 can be used to appreciate the effects of photometric contamination 
by a nearby, fainter star; the three entries for this star show a range in 
$M_{\rm V}$ of nearly one magnitude and distance variations of nearly a 
factor of two.

\section{Age comparisons}

Figure 10 shows a $c_{\rm 0},(b$--$y)_{\rm 0}$ diagram comparing the VMP stars, 
the metal-poor GC M92, and isochrones from the work of Bergbusch \& VandenBerg 
(2001).  For the VMP stars, the dereddened photometry and classifications are 
taken from Table 5.   The $uvby$ data for M92 is that of Grundahl et al.~(2000a), 
as provided by Grundahl (2000), has been corrected for a reddening of 
$E(b-y) = 0\fm016$, as suggested by these authors, and this CCD data has been 
plotted only for those stars with more than eight observations in the 
$u$--band and abs(SHARP) $\leq$ 0.05, from the DAOPHOT reduction package
(Stetson 1987); this latter parameter measures the goodness of fit 
between the PSF of the object and the model PSF and is used to exclude
non-stellar objects, double stars and stars affected by cosmic rays.

Also, the CCD $uvby$ data for M92 has been shifted by $-0\fm03$ in 
$c_{\rm 0}$, slightly less than the correction suggested by Grundahl et 
al.~(2000a).  They compared their $uvby$ data for M92 to that of local 
metal-poor stars from SN, especially the Hipparcos stars HD84937 and HD140283, 
using the $c_{\rm 0},(v-y)_{\rm 0}$ and M$_{\rm V},(v-y)_{\rm 0}$ diagrams, 
and concluded that their $c_{\rm 0}$ values should be corrected by about 
$\approx -0\fm04$; they suspect that this problem is due to a $u$--band 
zero-point error.  Indeed, in the $c_{\rm 0},(b$--$y)_{\rm 0}$ diagrams to 
follow (Figs.~10 and 11) we have noted a better overlap of the TO and SG 
distributions in $c_{\rm 0}$ if the M92 data is shifted downward by 
0\fm02--0\fm03, slightly less than that recommended by Grundahl et al.~(2000a). 

This is surprising since the $uvby$ data of our VMP stars and that of 
the M92 stars should both be closely on the same photometric system, that of 
Olsen (1983, 1984), which is also that of SN.  For the present catalogues the 
photometric standard stars were selected as described above, from Olsen 
(1983, 1984), from SN, and from S96.  SN took great care to transform their 
$uvby$ data onto the system of Olsen (1983, 1984), and for the S96 catalogue 
the photometric standards were taken from Olsen and from SN.  Grundahl et 
al.~(2000a) also selected their 55 $uvby$ standard stars used to calibrate the 
M92 data from Olsen (1983, 1984) and from SN.  So it all returns to that 
stated at the beginning of this paragraph:  both sets of data (VMP and M92) 
should be closely on the standard $uvby$ system defined by Olsen (1983, 1984).  

Plausibly, there might also be a shift in the relative $(b-y)$ systems as large
as $0\fm01$ between the data of Grundahl et al.~(2000a) and the present paper.
However, the $(b-y)$ observations are usually the easiest to transform to the
standard system of all the $uvby$ colors and indices, and these transformations
are linear over a wide color range (Gr{\o}nbech et al.~1976).  For the 
photoelectric observations, typical instrumental errors in $(b-y)$ for the
standard stars are $\pm0\fm003$, and typical transformation dispersions,
$\pm0\fm005$.  Systematic problems should be of this order or less.

The reddening value for M92 taken from Grundahl et al.~(2000a) is the canonical
value, and seems to be very well determined (Harris 1996, 2003; Schlegel et
al.~1998).  However, some authors (e.g.~King et al.~1998) have argued for a much
higher reddening for M92 ($0\fm09$--$0\fm10$) using indirect spectroscopic 
comparisons.  Such a high reddening for M92 seems to be highly unlikely
(VandenBerg 2000), but a possible error of $\pm 0\fm01$ in the canonical value
cannot be ruled out, and should be kept in mind during the following 
relative-age comparisons.

In Fig.~10 clear evidence can be seen that the youngest VMP stars are 
somewhat younger than the GC M92.  The upper extension along the 
isochrones of the VMP TO stars (the open squares) would indicate an age of 
12--13 Gyrs, while the upper extent of M92, about 14.0--14.5 Gyrs for a 
difference of $\approx 1.5$--2.0 Gyrs.  If one considers the transitional 
stars classified as BS--TO also as legitimate VMP turnoff stars, then 
31066--027, plotted as a filled diamond with the values 
($c_{\rm 0}, (b$--$y)_{\rm 0}$, [Fe/H]) = (0\fm431, 0\fm262, $-2.11$),  
respectively, plus a couple of little-evolved TO stars, would indicate an 
even larger range of $\ga 2.0$--2.5 Gyrs between M92 and the youngest VMP 
stars.  Our age for M92 agrees well with that derived by Grundahl 
et al.~(2000a), 14.5 Gyrs, but we do not agree with them that ``... the extremely 
metal-deficient field halo stars are most likely coeval with M92 to within 
1 Gyr.''  Other authors, such as Bell (1988), have also found that the bluest 
M92 stars are redder than the bluest VMP field subdwarfs, such as HD84937, by
about 0\fm03 in $(B-V)_{\rm 0}$, corresponding to 0\fm02 in $(b-y)_{\rm 0}$ and
an age difference of $\approx 2.5$ Gyrs, assuming similar metallicities.

However, these comparisons depend critically upon the [Fe/H] values used 
for the VMP field stars and for M92, i.e. upon the consistency between the 
[Fe/H] scale for field subdwarfs and VMP subgiants and the scale for GCs
in which the metallicities have been measured mainly for the
brighter red-giant stars.  For example, Bergbusch \& VandenBerg (2001) 
suggest that indeed there is an inconsistency between subdwarf and GC
[Fe/H] scales based upon their fitting of isochrones to observed 
color-magnitude diagrams for
GCs.  More specifically, the above value of [Fe/H] $= -2.22$ for M92 has
been obtained by Grundahl et al.~(2000a) using sources based upon the
high-resolution spectroscopy of the brighter red-giant stars and upon the
calibration of the integrated light of GCs, while King et al.~(1998)
obtained [Fe/H] $= -2.52$ for M92 from high-resolution spectroscopic
observations of three subgiants, a factor of two lower for the iron to
hydrogen ratio.  If this latter metallicity is indeed the correct one for
M92, then in Fig.~10, VMP field stars with [Fe/H] $\approx -2.22$ are being 
compared to a GC with [Fe/H] $= -2.52$.  According to the isochrones of 
Bergbusch \& VandenBerg (2001), as transformed to $uvby$ by Clem et al.~(2003), 
a correction for this metallicity difference would increase the 
age differences discussed above by about 1.5 Gyrs.  A more recent study of
the GC [Fe/H] scale by Kraft \& Ivans (2003) suggests that at least part
of the inconsistency with the subdwarf scale is due to non-LTE 
``overionization'' effects for Fe I lines.  For six red giants in M92 they 
obtain an average [Fe/H] = $-2.38$ from an analysis of Fe II lines only.
Such a metallicity for M92 would require a correction of +0.8 Gyr 
to the age differences discussed above for Fig.~10.

In Fig.~11 the comparison of Fig.~10 is repeated, but now with the field
VMP stars drawn from the range $-2.67<$ [Fe/H] $<-2.37$, which is centered on
the value [Fe/H] $= -2.52$ for M92 obtained by King et al.~(1998).  This 
comparison would indicate age differences not that distinct from those of 
Fig.~10, despite the change in the mean metallicity of the field stars.
Three TO stars along the axis of the isochrones would again suggest that the
youngest VMP stars are 1.0--1.5 Gyrs younger than M92.  The BS--TO star
22876--039 (0\fm486, 0\fm256, $-2.60$), and two little-evolved TO stars 
would indicate larger age differences, $\ga 3.0$ Gyrs.

These results are somewhat surprising, considering that M92 is among the
more metal-poor and older GCs of the Galaxy (VandenBerg 2000), and that there
is evidence that the formation of all metal-poor Galactic GCs was triggered
throughout the Galaxy at the same time to within $\approx 1$ Gyr (Harris et al.~1997; 
Lee et al.~2001).  Also, several previous studies (such as those of 
Pont et al.~1998 and of Grundahl et al.~2000a) have concluded that the more 
metal-poor field subdwarfs are coeval with M92 to within about 1 Gyr; 
however, these works have in general used only the more local subdwarfs, 
such as those from SN or Hipparcos.  The VMP stars of this paper span a 
larger volume in the Galaxy, and the younger VMP stars of Figs.~10 and 11,
which appear to be at least 1--3 Gyrs younger than M92, may reveal evidence for 
the belated formation of VMP stars outside of the Galactic GCs, 
the hierarchical infall of VMP material from the outermost parts of 
the proto-Galaxy after the GC system had formed (Sandage 1990), and/or the 
accretion of material from another galaxy with formation and 
chemical-enrichment histories different from that of the Galaxy (Preston 
et al.~1994; Ibata et al.~1994).  For example, Preston et al.~have concluded
that their blue metal-poor stars ([Fe/H] $< -1.0$ and $0\fm15 < (B-V)_{\rm 0}
< 0\fm35$, bluer than the GC turnoffs) are probably the result of accretion
events by the Galaxy of material from dwarf galaxies, and the study of
seven dwarf spheroidals by Dolphin (2002) has indeed shown recent (0.5--5 Gyrs) star
formation for more than half of these (Carina, Leo I, Leo II, and Sagittarius),
but higher metallicities ([Fe/H] $\approx -1.0$ to $-1.2$) than the present
VMP stars.  However, a previous compilation by Mateo (1998) gave [Fe/H] $\approx
-1.9$ to $-2.0$ for Carina and Leo II, more in line with our VMP stars, but
requiring an increase in the age estimates of Dolphin by $\approx 5$ Gyrs. 
Nevertheless, stars with [Fe/H] $\approx -2.0$ and ages of 5--10 Gyrs would come 
close to explaining the bluer VMP TO stars of Figs.~10 and 11.


   \begin{figure}
   \centering
   \includegraphics[width=12cm]{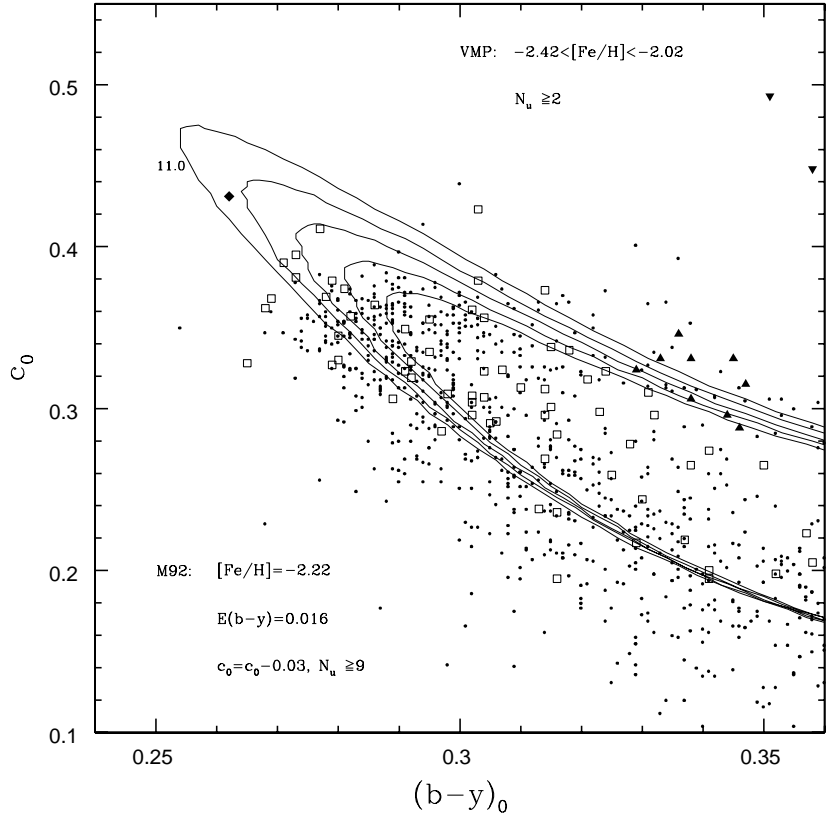}
      \caption{Comparison in the $c_{\rm 0},(b$--$y)_{\rm 0}$ 
               diagram of the VMP 
               stars with $-2.42 \leq$ [Fe/H] $\leq -2.02$ (with the 
               same symbols as defined in Fig.~6), with the $uvby$ 
               observations of M92 provided by Grundahl (2000a; 
               points), and with the isochrones of Bergbusch \& 
               VandenBerg (2001) as transformed to $uvby$ 
               photometry by Clem et al.~(2003; solid curves). 
               The isochrones are those for [Fe/H] $= -2.31$ and 
               [$\alpha$/Fe] $=+0.30$, and have been plotted for 11,
               12, 13, 14, and 15 Gyrs.  The [Fe/H] of M92 has been 
               taken here to be $-2.22$ according to the average
               derived by Grundahl et al.~(2000a) from several high
               resolution spectroscopic studies of cluster giants
               found in the literature.  Only those VMP stars with
               two or more $uvby$ observations have been plotted,
               and only those GC stars with more than eight $u$
               observations and abs(SHARP) $\leq 0.05$.  The M92
               photometry has been shifted by $-0\fm03$ in $c_{\rm 0}$,
               similar to that correction suggested by Grundahl
               et al.~(2000a). 
              }
         \label{FigM92m2.22Isos}
   \end{figure}
%

   \begin{figure}
   \centering
   \includegraphics[width=12cm]{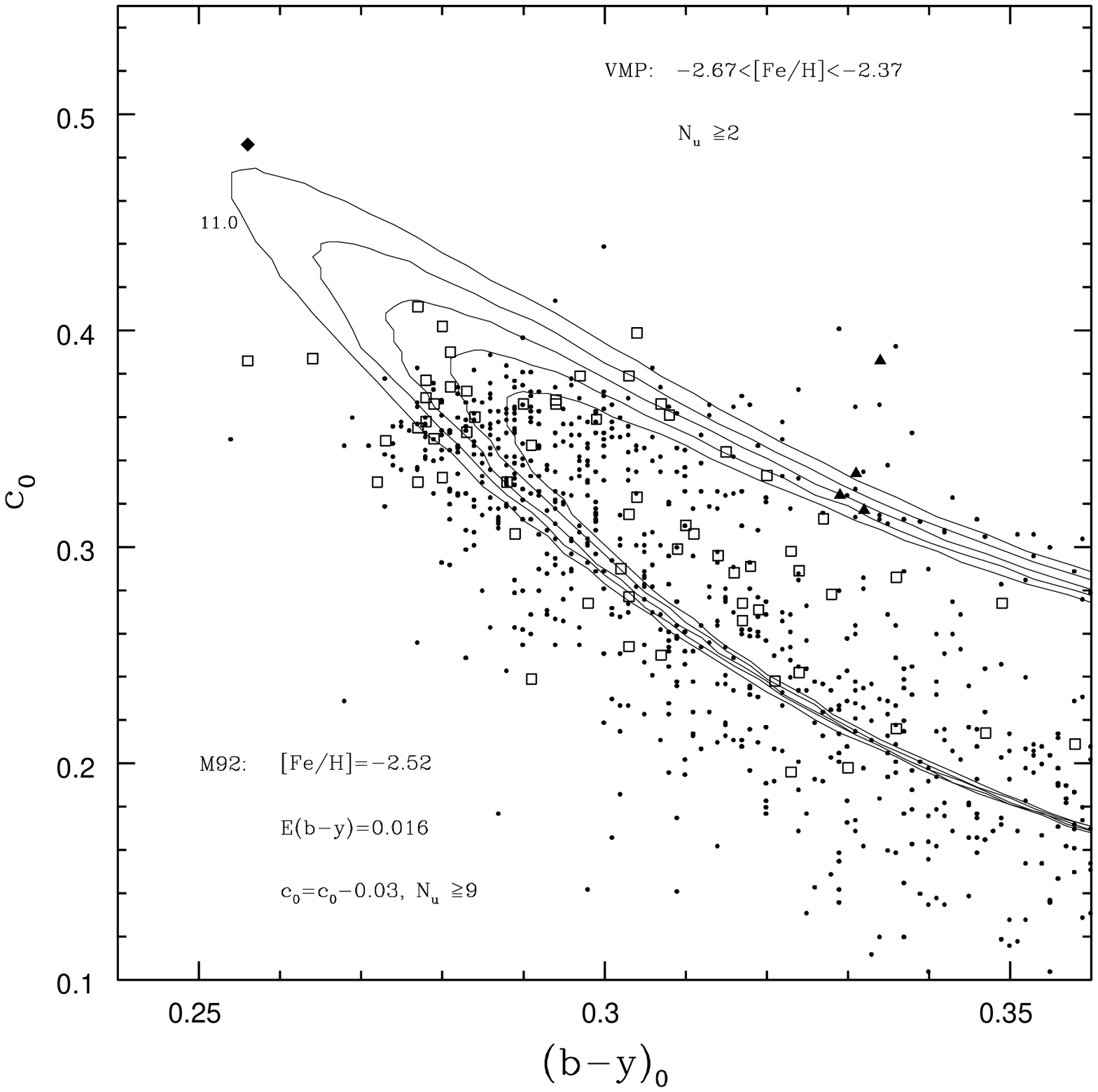}
      \caption{The same $c_{\rm 0},(b$--$y)_{\rm 0}$ diagram as Fig.~10 
               but assuming 
               that M92 has [Fe/H] $= -2.52$ according to the study of
               King et al.~(1998) from high-resolution spectroscopy
               of three subgiants.  The VMP stars are plotted for the
               range $-2.67 \leq$ [Fe/H] $\leq -2.37$ (again with the 
               same symbols as Fig.~6), with the same $uvby$ 
               observations of M92 and with the same isochrones as in
               Fig.~10.  Again, only those VMP stars with
               two or more $uvby$ observations have been plotted,
               and the M92 data has been shifted by $-0\fm03$ in 
               $c_{\rm 0}$.  Isochrones have been plotted for 11, 12,
               13, 14, and 15 Gyrs.
               }
         \label{FigM92m2.52Isos}
   \end{figure}
%
%

\section{Conclusions}

   \begin{enumerate}
      \item The overall VMP HK-survey sample contains a wide range of stellar
         types, ranging from horizontal branch stars to subluminous, and from 
         red giant stars to the blue horizontal branch.
      \item The dereddened $c_{\rm 0},(b-y)_{\rm 0}$ diagram has been shown to
         be quite useful for providing photometric classifications of the VMP
         stars analogous to types derived from GC color-magnitude diagrams, 
         such as Turn-Off stars (TO), SubGiants (SG), Red Giants (RG), Horizontal 
         Branch stars (HB), Blue Horizontal Branch stars (BHB), Blue Stragglers (BS), 
         SubLuminous stars (SL), and so forth (see Fig.~6). 
      \item The intrinsic-color calibration of Schuster \& Nissen (1989), as
         modified slightly by Nissen (1994), is shown to provide reddening
         excesses, $E(b-y)$ or $E(B-V)$, very similar to the adopted reddening
         estimates derived in this publication from the maps of Schlegel,
         Finkbeiner, \& Davis (1998) (see Eq.~1).  No significant systematic
         offsets between these two dereddening techniques are noted (see Fig.~3).
      \item A number of VMP stars have been noted with probable anomalous
         photometric traits, especially from the $m_{\rm 1}$ and $[m_{\rm 1}]$
         indices; two such groups stand out.  First, there are several stars 
         with $(b$--$y)_{\rm 0} \la 0\fm45$ and with $m_{\rm 0} \ga 0\fm17$, 
         much larger than would be expected for VMP stars with 
         [Fe/H] $\la -1.5$.  Most of these have been classified SG, and
         some show clear evidence of photometric variability.  These are 
         perhaps analogous to stars discussed in S96 with larger than
         expected $[m_{\rm 1}]$ values.  We suggest here that these are
         misclassified AGB stars with unusual chemical-abundance ratios,
         photometric variability, and/or binary companions.
      \item The second group of anomalous stars are those ten classified BS
         and having $m_{\rm 1}$, $[m_{\rm 1}]$, and $(U-B)_{\rm 0}$ values
         indicating nearly solar [Fe/H] values.  There is a clear discrepancy
         here between these photometric indices and the KP index used to derive
         [Fe/H] for the HK survey.  These stars are very similar to the Am
         stars identified by Wilhelm et al.~(1999a, 1999b) and have been
         noted as ``BS (Am)'' in Table 5.
      \item The photometric distances from the $UBV$ and $uvby$ photometries 
         agree reasonably well considering the problems, lack of calibrating 
         stars, and extrapolations needed for the more VMP stars.  Our 
         Hipparcos-based, photometric calibration for $M_{\rm v}$ seems to 
         work quite well for the turn-off, main-sequence, and subgiant VMP 
         stars, as suggested in Figs.~7 and 8.
      \item In the $c_{\rm 0},(b-y)_{\rm 0}$ diagram, the youngest VMP stars
         appear to have ages 1--3 Gyrs younger than the GC M92.  Uncertainties
         in the [Fe/H] scale for M92 would tend to increase this age difference 
         even more.  (The interstellar reddening of M92 seems to be well determined
         but might be as uncertain as $\pm 0\fm01$).  Such younger VMP stars are 
         showing evidence for important details upon the overall formation and 
         evolution of the Galaxy, such as possible hierarchical 
         star-formation/mass-infall for the VMP material, and/or accretion 
         processes from other (dwarf) galaxies with different formation and 
         chemical-enrichment histories.
      
      \end{enumerate}

\begin{acknowledgements}
      One of us (W.J.S.) is very grateful to UNESCO (the United Nations
      Development Programme) for funding which supported travel to Chile for 
      one of the observing runs at La Silla, Chile, to the DGAPA--PAPIIT (UNAM) 
      (projects Nos. IN101495 and IN111500) and to CONACyT (M\'exico) 
      (projects Nos. 1219--E9203 and 27884E) for funding
      which permitted travel and also the maintenance and upgrading of 
      the $uvby$--$\beta$ photometer, and also to Jos\'e Guichard, who extended
      the invitation to spend my sabbatical year at INAOE, where much of the
      text for this publication has been written and the final versions of most of
      the tables and figures prepared.  T.C.B. acknowledges partial support 
      for this work from grants AST 00-98508 and AST 00-98549 awarded by the 
      U.S. National Science Foundation. Much of this paper would not have 
      been possible without the GC $uvby$ data provided by Frank Grundahl; 
      we are extremely grateful!  Chris 
      Flynn helped with the observations during a couple of nights at the
      beginning of the October 1998 run at La Silla, his assistance made our
      observing more efficient, and we greatly appreciate it.  Don VandenBerg
      and James Clem made their $uvby$ isochrones available prior to
      publication, and we are very grateful.  We also thank greatly Laura
      Parrao who helped with part of the data reductions of the $uvby$--$\beta$
      photometry taken in M\'exico at the National Astronomical Observatory
      and who also helped with some of the calibrations and analyses.  We thank A. 
      Franco and S. Ruiz-Berbena, who helped with some of the preliminary 
      analyses.  Many people at the SPM observatory have helped over the years 
      with the maintenance and upgrading of the photometer, with computer 
      programing and debugging, and with library sources; we thank especially 
      L. Gut\'{\i}errez, V. Garc\'{\i}a (deceased), B. Hern\'andez, J.M. Murillo, 
      J.L. Ochoa, J. Valdez, B. Garc\'{\i}a, B. Mart\'{\i}nez, E. L\'opez, 
      M.E. Jim\'enez, and G. Puig.  Last but not least, we would like to thank
      N. Christlieb, the referee, for a very careful reading of the manuscript
      and several useful changes to the style and presentation.
         
\end{acknowledgements}

\end{document}